\documentclass[10pt]{wlscirep}
\usepackage[utf8]{inputenc}
\usepackage{setspace}
\usepackage{parskip}
\usepackage{amssymb}
\usepackage{amsmath}
\usepackage{graphicx}
\usepackage[english]{babel}
\usepackage{fancyhdr}
\usepackage[T1]{fontenc}
\usepackage{gensymb}
\usepackage{epstopdf}
\usepackage{siunitx}
\usepackage{graphicx} 
\usepackage{float}
\usepackage{cite}
\usepackage{wasysym}
\usepackage{mathtools}
\usepackage{authblk}
\usepackage{titlesec}
\usepackage[numbers,square,super]{natbib}
\usepackage[export]{adjustbox}
\usepackage{textgreek}

\titlelabel{\thetitle.\quad}
\makeatletter
\newcommand\bigDiamond{\mathop{\mathpalette\bigDi@mond\relax}}
\newcommand\bigDi@mond[2]{%
  \vcenter{\hbox{\m@th
    \scalebox{\ifx#1\displaystyle 2\else1.2\fi}{$#1\Diamond$}%
  }}%
}
\newcommand\bigLozenge{\mathop{\mathpalette\bigL@zenge\relax}}
\newcommand\bigL@zenge[2]{%
  \vcenter{\hbox{\m@th
    \scalebox{\ifx#1\displaystyle 2\else1.2\fi}{$#1\blacklozenge$}%
  }}%
}
\makeatother

\begin{document}

\title{Thermoplasmonic response of semiconductor nanoparticles: A comparison with metals}

\author[1,*]{Vaibhav Thakore}
\author[2,\textdaggerdbl]{Janika Tang}
\author[2,\dag]{Kevin Conley}
\author[2,3,4,\S]{Tapio Ala-Nissila}
\author[1,5,\$]{Mikko Karttunen}

\affil[1]{Department of Applied Mathematics, Western University, 1151 Richmond Street, London, Ontario N6A 5B7, Canada}
\affil[2]{QTF Center of Excellence, Department of Applied Physics, Aalto University School of Science, FIN-00076, Aalto, Espoo, Finland}
\affil[3]{Department of Physics, Brown University, Providence, Rhode Island 02912-1843, USA}
\affil[4]{Interdisciplinary Centre for Mathematical Modelling, Department of Mathematical Sciences, Loughborough University, Loughborough LE11 3TU, UK}
\affil[5]{Department of Chemistry, Western University, 1151 Richmond Street, London, Ontario N6A 5B7, Canada}

\affil[*]{Email: vthakore@knights.ucf.edu,
	      Phone: +1 226 239 4731}
\affil[$\ddag$]{janika.tang@aalto.fi}
\affil[$\dag$]{kevin.conley@aalto.fi}
\affil[$\S$]{tapio.ala-nissila@aalto.fi}
\affil[$\$$]{mkarttu@uwo.ca}

\begin{abstract}
A number of applications in nanoplasmonics utilize noble metals, gold (Au) and silver (Ag), as the materials of choice. However, these materials suffer from problems of poor thermal and chemical stability accompanied by significant dissipative losses under high-temperature conditions. In this regard, semiconductor nanoparticles have attracted attention with their promising characteristics of highly tunable plasmonic resonances, low ohmic losses and greater thermochemical stability. Here, we investigate the size-dependent thermoplasmonic properties of semiconducting silicon and gallium arsenide nanoparticles to compare them with metallic Au nanoparticles using Mie theory. To this end, we employ experimentally estimated models of dielectric permittivity in our computations. Among the various permittivity models for Au, we further compare the Drude-Lorentz (DL) and the Drude and critical points (DCP) models. Results show a redshift in the scattering and absorption resonances for the DL model while the DCP model presents a blueshift. The dissipative damping in the semiconductor nanoparticles is strongest for the sharp electric octupole resonances followed by the quadrupole and dipole modes. However, a reverse order with strongest values for the broad dipole resonance is observed for the Au nanoparticles. A massive Drude broadening contributes strongly to the damping of resonances in Au nanoparticles at elevated temperatures. In contrast, the semiconductor nanoparticles do not exhibit any significant deterioration in their scattering and absorption resonances at high temperatures. In combination with low dissipative damping, this makes the semiconductor nanoparticles better suited for high-temperature applications in nanoplasmonics wherein the noble metals suffer from excessive heating.  
\end{abstract}

\flushbottom
\lhead{Thermoplasmonic response of semiconductor nanoparticles}
\maketitle

\section*{INTRODUCTION}
Plasmonically enhanced absorption or scattering of radiation at the mesoscale forms the basis of promising applications in a wide variety of fields. Applications as diverse as: biosensing for lab-on-chip devices \cite{RN142}; photothermal therapy in cancer treatment \cite{RN31,RN32}; energy-efficient spatiotemporal control of photocatalysis \cite{RN144}; solvothermal chemistry for radiation-controlled synthesis of materials \cite{RN143}; energy harvesting using solar/thermophotovoltaics and thermoelectrics \cite{RN124}; heat-assisted magnetic recording for high density data storage \cite{RN136}; and, control of radiative heat transfer in insulators \cite{RN135} - all exploit the plasmonic behavior of nano- or microstructured materials. Also, nearly all of these applications either lead to generation of heat due to losses in the plasmonic materials/devices or require their operation at elevated temperatures. For many applications based on plasmonics the noble metals - gold (Au) and silver (Ag) - have been the materials of choice. However, it is also common knowledge that these materials are characterized by problems of poor thermal and chemical stability due to large ohmic dissipation at high temperatures \cite{RN145, RN137}. Furthermore, the tunability of the plasmonic resonances in the noble metals is limited to the ultraviolet and visible frequency regimes of electromagnetic radiation \cite{RN62}. These issues have prompted a quest for materials with better thermoplasmonic properties for applications that require operating in harsh environments \cite{RN126}. The use of thermochemically stable semiconducting materials has thus assumed importance since they present low dissipative damping and highly tunable plasmonic resonances through bandgap engineering and control over dopant concentration \cite{RN62, RN89}. The widespread use of noble metals in applications based on plasmonics has also meant that their thermoplasmonic properties have been extensively studied and are well-characterized \cite{RN125, RN114, RN122}. However, despite their obvious technological relevance, there have not been any studies reported, to our knowledge, on the thermoplasmonic properties of semiconducting nanoparticles.  

A large proportion of the energy absorbed by a plasmonic particle from the incident radiation is eventually either dissipated as heat or scattered away. The particle size and material dependent strength of the electron-electron (\textit{e-e}), electron-phonon (\textit{e-$\phi$}) interactions and the magnitude of the surface scattering of the charge carriers determine the contribution of the dissipative damping to the portion of the absorbed energy that goes into generating heat. This dissipation of the absorbed energy as heat thus leads to a rise in the temperature of the nanoparticle. Depending on the material and the size of the nanoparticle, the resulting rise in temperature may cause a further increase in the absorption of the incident radiation and consequently its temperature through a nonlinear feedback effect \cite{RN147, RN36}. Additionally, there is a material, particle size and temperature dependent emission of photons by the plasmonic excitations referred to as radiation damping that leads to a decay of the absorbed energy through scattering. Such plasmonically enhanced scattering serves to minimize the temperature increase due to dissipative damping and thus helps cool the nanoparticle \cite{RN150}. It is therefore important to understand the strength and the nature of these two competing damping mechanisms, both for making an informed choice of a material for a target application and for the tailoring of its desired thermoplasmonic behavior.

Here, we report results from a computational study based on the Mie theory for size-dependent thermoplasmonic properties of undoped silicon (Si) and gallium arsenide (GaAs) with indirect and direct bandgaps, respectively. We further compare and contrast our results from spherical Si and GaAs nanoparticles with those obtained from temperature dependent plasmonic behavior of Au nanoparticles. Our results indicate that there is no significant deterioration in the scattering and absorption resonances at elevated temperatures for the semiconductor nanoparticles as compared to the metallic Au nanoparticles. Furthermore, the Au nanoparticles exhibit increased Drude damping of plasmonic resonances and enhanced absorption at longer wavelengths that is responsible for excessive heating at high temperatures. The theory and methods section presents the details of the dielectric permittivity models employed for computing the temperature dependent complex refractive indices of Si, GaAs and Au followed by the relevant details from the Mie theory.  

\section*{THEORY AND METHODS}
The thermoplasmonic response of a material is a function of its frequency and temperature dependent complex dielectric permittivity. The dielectric permittivity, in turn, is modulated by the changes in the scattering of the charge carriers and their concentration based on the energy of the incident photons. The scattering of the charge carriers has contributions from collisions among themselves due to Coulombic interactions and from collisions with phonons that typically increase in density with a rise in the system temperature \cite{RN114, RN118}. In confined systems, such as a nanoparticle, another contribution to scattering comes from the boundaries or surfaces \cite{RN146} that becomes significant for metallic nanoparticles with radii less than $\sim$$10$ nm. In reality, however, the system boundaries are not always smooth and one needs to additionally account for the effect of the surface roughness as well  \cite{RN131}. In even smaller nanoclusters (radii$\sim$$1$-$5$ nm), the number of charge carriers for both metallic and semiconducting systems becomes really small and the discreteness of the band-structure needs to be taken into account to model dielectric permittivity and plasmonic response \cite{RN116, RN30, RN29, RN121}.  However, to understand and isolate the differences in the thermoplasmonic response of semiconducting and metallic nanoparticles, we limit ourselves to the temperature dependent response of ideal spherically smooth particles whose behavior can be modeled using the Mie theory \cite{RN102, RN135}. We further include nanoparticles that can be treated as large enough to exhibit a continuous band structure. In addition, we employ \textit{experimentally estimated realistic models} of the dielectric permittivity of Au, Si and GaAs to compute their thermoplasmonic response \cite{RN114, RN117, RN115}.

\subsection*{Dielectric permittivity}
The temperature dependent refractive indices of Au and the intrinsic semiconductors (Si and GaAs) required for the computation of the particle size-dependent thermoplasmonic response are computed in the wavelength ranges $\lambda = 400$-$1450$ nm and $500$-$1450$ nm respectively using the following methodologies. 
\subsubsection*{Gold}
For modeling the dielectric permittivity of metals either a Drude-Lorentz (DL) or a Drude and critical points (DCP) model is employed. The Drude-Lorentz model can in principle include an arbitrary number of oscillator terms to approximate the spectral line shapes due to interband transitions with possibly no direct physical relevance to the optical response of a material \cite{RN147, RN114}. The Drude and critical points model on the other hand is causal and physically motivated as it takes into account the asymmetric line shapes due to interband transitions at the critical points \cite{RN114}. However, the use of the simple DCP model may also present problems \cite{RN125}. Therefore, we employ both the DL and DCP models to understand (a) the important differences between them, and, (b) their predictions of the thermoplasmonic behavior of Au particles. 

Briefly, for the DL model, we compute the dielectric permittivity for Au based on the model estimated by Rakic \textit{et al.} and given by \cite{RN118}
\begin{equation}
\epsilon(\omega) = \epsilon_{\infty} - \frac{\Omega_{\mathrm{p}}^{2}}{\omega^{2}+i\Gamma_{\mathrm{D}}\omega} + \sum^5_{j=1} \frac{C_j\omega_{\mathrm{p}}^{2}}{(\omega_j^2 - \omega^2) + i\omega\gamma_j}.
\label{epsilonDL}
\end{equation}
Here, $\Omega_\mathrm{p} = \sqrt{f_\mathrm{0}}\omega_\mathrm{p}$ is the plasma frequency associated with the intraband transitions of oscillator strength $f_\mathrm{0}$ and the Drude damping factor $\Gamma_\mathrm{D}$; $\epsilon_{\infty}$, $\omega$ and $\omega_{\mathrm{p}}$ are the background dielectric constant, frequency of the incident radiation and the plasma frequency, respectively. The parameters $C_j$, $\omega_j$, and $\gamma_j$ represent the oscillator strength, characteristic frequency and damping respectively for the five Lorentz oscillators considered in the DL model \cite{RN118}. The temperature dependence of the permittivity in the DL model comes from the second term in equation \eqref{epsilonDL} through the plasma frequency $\Omega_\mathrm{p}$ and the Drude damping factor $\Gamma_\mathrm{D}$. However, the Lorentz oscillator parameters $C_j$, $\omega_j$ and $\gamma_j$ in the DL model, as estimated by Rakic \textit{et al.} from experimental ellipsometric data, are assumed to be independent of the temperature \cite{RN118}. The temperature dependence of the plasma frequency $\omega_\mathrm{p}$ [$=Ne^2/m^*\epsilon_\mathrm{0}$] is related to the changes in the carrier concentration $N$ under the influence of the lattice thermal expansion and is given by \cite{RN147}
\begin{equation}
\omega_\mathrm{p}(T) = \frac{\omega_\mathrm{p}(T_\mathrm{0})}{\sqrt{1 + 3\alpha_\mathcal{L}(T - T_\mathrm{0})}}.
\label{omegaT}   
\end{equation}
Here, $\alpha_\mathcal{L}$ is the coefficient of linear thermal expansion for Au, $T$ is the temperature, $T_\mathrm{0} = 293.15$ K is the reference temperature, $m^*$ is the effective mass of the electrons, $e$ is the electric charge and $\epsilon_\mathrm{0}$ is the permittivity of the free space. The Drude damping $\Gamma_\mathrm{D}$ ($= \Gamma_\mathrm{ee} + \Gamma_\mathrm{e\phi}$) has temperature dependent contributions from both electron-electron (\textit{e-e}) and electron-phonon (\textit{e-$\phi$}) interactions through the corresponding damping factors $\Gamma_\mathrm{ee}$ and $\Gamma_\mathrm{e\phi}$ that are given by \cite{RN147, RN132, RN119}
\begin{equation}
\Gamma_\mathrm{ee} = \frac{\pi^3\Gamma\Delta}{12E_\mathrm{F}}\bigg[(k_\mathrm{B}T)^2 + \bigg(\frac{\hslash\omega}{2\pi}\bigg)^2\bigg]
\label{g_ee}
\end{equation}
and
\begin{equation}
\Gamma_\mathrm{e\phi} = \Gamma_\mathrm{o}\bigg[\frac{2}{5} + 4\bigg(\frac{T}{\Theta_\mathrm{D}}\bigg)^5 \int^{\Theta_\mathrm{D}/T}_0\frac{z^4}{e^z - 1}\mathrm{d}z\bigg]. 
\label{g_eph}
\end{equation}
Here, $\Gamma=0.55$ is a constant determining the average scattering probability over the Fermi surface; $\Delta = 0.77$ is the fractional umklapp scattering; $E_\mathrm{F} = 5.51$ eV is the Fermi energy of the free electrons in Au; $\Theta_\mathrm{D} = 177$ K is the Debye temperature for Au; and, $\Gamma_\mathrm{o} = 0.07$ eV is a material dependent constant obtained by fitting the bulk permittivity of Au for energies of incident radiation less than those corresponding to the onset of the interband transitions \cite{RN147, RN132, RN119}.

For the DCP model, we employ the experimentally estimated model by Reddy \textit{et al.} with dielectric permittivity of the form \cite{RN114}
\begin{equation}
\epsilon(\omega) = \epsilon_{\infty} - \frac{\omega_{\mathrm{p}}^{2}}{\omega^{2}+i\Gamma_{\mathrm{D}}\omega} + \sum^2_{j=1} C_{j}\omega_{j}\bigg[\frac{e^{i\phi_{j}}}{\omega_{j} -\omega - i\gamma_{j}} + \frac{e^{-i\phi_{j}}}{\omega_{j} + \omega + i\gamma_{j}}\bigg]
\label{epsilonDCP}
\end{equation}
where the parameter $\phi_j$ represents the oscillator phase for the two oscillators in the DCP model. The rest of the critical point oscillator parameters $C_j$, $\omega_j$ and $\gamma_j$ have the same meaning as for the Lorentz oscillators in the DL model. The critical point oscillators correspond to the two interband transitions observed in Au around $\sim$$470$ and $\sim$$330$ nm \cite{RN151}. Reddy \textit{et al.} estimate the temperature dependent DCP model parameters for a number of crystalline and polycrystalline Au films of different thicknesses. These films present different levels of thermophysical stability during the measurement of their optical spectra between $23$-$500$ \degree C \cite{RN114}. However, for our study, we employ the DCP model parameters estimated for the 200 nm thick single-crystalline Au film that was observed to be stable at all temperatures under repeated cycles of measurement \cite{RN114}. 

\subsubsection*{Silicon}
The temperature dependence of the refractive index $n_p$ for Si is computed using an empirical power law   
\begin{equation}
\zeta(T) = \zeta(T_\mathrm{0})(T/T_\mathrm{0})^{b}
\end{equation}
proposed by Svantesson \textit{et al.} \cite{RN138} and employed by Green to model the dielectric dispersion in intrinsic Si \cite{RN117}. Here, $\zeta$ is the real ($\eta$) or the imaginary ($\kappa$) part of the refractive index $n_\mathrm{p}$, $T$ and $T_\mathrm{0} = 300$ K are the temperature and the reference temperature, respectively, and the exponent $b$ is related to the normalized temperature coefficients $C_{\zeta}$ [$\equiv (1/\zeta)d\zeta/dT$] of the optical constants $\eta$ and $\kappa$ through the expression
\begin{equation}
b = C_{\zeta}(T)T = C_{\zeta}(T_\mathrm{0})T_\mathrm{0}.
\end{equation}
The optical constants $\eta$ and $\kappa$ for Si are themselves estimated using the accurate spectroscopic ellipsometry data reported by Herzinger \textit{et al.} \cite{RN148} and consistently reconciled with the data from previous researchers by Green \cite{RN117}. The calculations for the optical constants are based on a Kramers-Kronig analysis \cite{RN152} that makes use of the reflectance $\mathcal{R}$ to first compute its phase $\theta$ in radians at the target energy $E_t$ as 
\begin{equation}
\theta(E_{\mathrm{t}}) = -\frac{E_{\mathrm{t}}}{\pi}\bigg[\int_{0}^{E_{\mathrm{t}} - \delta}\frac{\ln[\mathcal{R}(E)]}{E^2 - E_{t}^{2}}dE + \int_{E_{\mathrm{t}} - \delta}^{\infty}\frac{\ln[\mathcal{R}(E)]}{E^2 - E_{t}^{2}}dE\bigg]
\end{equation}
where $\delta\rightarrow 0$ and $E$ is a dimensionless quantity. The real ($\eta$) and the imaginary ($\kappa$) parts of the refractive index are then calculated as 
\begin{equation}
\eta(E) = \frac{1 - \mathcal{R}(E)}{1 + \mathcal{R}(E) - \sqrt{2\mathcal{R}(E)}\cos\theta}
\end{equation}
\begin{equation}
\kappa(E) = \frac{2\sqrt{\mathcal{R}(E)}\sin\theta}{1 + \mathcal{R}(E) - \sqrt{2\mathcal{R}(E)}\cos\theta}.
\end{equation} 
The temperature dependent dielectric permittivity for Si is then given by 
\begin{equation}
\epsilon(T) = n_p^2(T) = [\eta(T) + i\kappa(T)]^2 = \eta^2 - \kappa^2 + 2i\eta\kappa.
\label{epsilonNK}
\end{equation}
\subsubsection*{GaAs}
We employ a heuristic model proposed and estimated by Reinhart for photon energies below, at and above the bandgap to model the temperature dependence of the refractive index for the direct bandgap intrinsic GaAs \cite{RN115}. This model realistically accounts for the existence of the experimentally observed band tails and the non-parabolic shape of absorption above the bandgap in undoped GaAs. The model requires estimating the continuum ($\alpha_\mathrm{c}$) and the exciton ($\alpha_\mathrm{ex}$) contributions to the absorption coefficient $\alpha$ $(=\alpha_\mathrm{c} + \alpha_\mathrm{ex})$ by fitting the following functions to the experimental absorption spectrum \cite{RN115}
\begin{equation}
\alpha_\mathrm{c}(E') = A\ \exp[r(E')]\bigg[\frac{1}{1 + \exp(-E'/E_\mathrm{s1})} + \frac{a_\mathrm{so}}{1 + \exp((\Delta-E')/E_\mathrm{so})}\bigg],
\label{alphaC}
\end{equation}
and
\begin{equation}
\alpha_\mathrm{ex}(E') = \sum_{j=1}^{2}\bigg[ \frac{a_{\mathrm{x}j}}{\exp[(E' + E_{\mathrm{x}j})/E_\mathrm{s1}] + \exp[-(E' + E_{\mathrm{x}j})/E_\mathrm{s2}]}\bigg],
\label{alphaE}
\end{equation} 
where
\begin{align*}
r(E') = r_{1}E' + r_{2}E'^{2} + r_{3}E'^{3}\ \mathrm{and}\ E' = E - E_\mathrm{g}(T).
\end{align*}
The first term in equation \eqref{alphaC} describes the increase in the absorption $\alpha$ above the fundamental bandgap while the Fermi-function-type terms represent the sharp step-like absorption increase at the fundamental and the split-off gaps. The temperature dependent bandgap-shrinkage effect is modeled using \cite{RN130}
\begin{equation}
E_\mathrm{g}(T) = E_\mathrm{g}(0) - \frac{\sigma \Theta}{2}\bigg[\sqrt[p]{1 + \bigg(\frac{2T}{\Theta}\bigg)^p}  - 1\bigg].
\end{equation}
For intrinsic GaAs: $E_\mathrm{g}(0) = 1.5192$ eV is the zero-temperature bandgap; $\sigma = 0.475$ eV/K is the high-temperature limiting value of the forbidden-gap entropy; $\Theta = 222.4$ K is the material specific phonon temperature that represents the effective phonon energy ($\hslash\omega_\mathrm{eff}\equiv \mathrm{k_B}\Theta$, $\hslash$ is the reduced Planck's constant); and, $p = 2.667$ is an empirical parameter that is a consequence of the globally concave shape ($p > 2$) of the electron-phonon spectral distribution function in conjunction with the contributions from the thermal expansion mechanism to the total gap-shrinkage effect \cite{RN130}.

The parameters $A$ and $a_\mathrm{so}$ in equation \eqref{alphaC} are the absorption amplitudes for the fundamental and the split-off valence band transitions respectively whereas the coefficients $r_j$ describe the increase in the absorption for  energies $E > E_\mathrm{g}$. The energy slope parameters $E_{\mathrm{s1}}$ and $E_\mathrm{s2}$ in equation \eqref{alphaE} along with $E_\mathrm{so}$ in equation \eqref{alphaC} for the split-off valence band are a linear function of the temperature that account for the weak electron-phonon coupling and are given by \cite{RN115}
\begin{equation}
E_\mathrm{s1} = 7(T+7)/304 \mathrm{,}\quad E_\mathrm{s2} = 4(T+7)/304 \mathrm{\quad and,}\quad E_\mathrm{so} = 40(T+7)/304\quad \mathrm{(in\ meV)}.
\end{equation}
The temperature dependent discrete transition amplitudes $a_{\mathrm{x}j} (j = 1,2)$ in equation \eqref{alphaE} relate to the transitions of the excitonic ground and first excited states. These are computed as   \cite{RN115}
\begin{equation}
a_{\mathrm{x1}} = a_{\mathrm{x0}}\ \exp(-T/230)[1 - \exp(-E_\mathrm{x1}/k_\mathrm{B}T)]
\end{equation}
and
\begin{equation}
a_{\mathrm{x2}} = a_{\mathrm{x0}}\ \exp(-T/230)[1 - \exp(-E_\mathrm{x2}/k_\mathrm{B}T)]/8.
\end{equation}
The exciton binding energies of the fundamental $(E_\mathrm{x1})$ and the first excited $(E_\mathrm{x2})$ states are $E_\mathrm{x1} = 4E_\mathrm{x2} = 4.22$ meV. 

The imaginary part $\kappa$ of the refractive index $n$ can thus be calculated from the absorption coefficient $\alpha$ as a function of the energy ($\mathcal{E}$) using the Kramers-Kronig integral as  \cite{RN115, RN152} 
\begin{equation}
\kappa(E) = \frac{c \hslash}{q\pi} \int_{0}^{E_\mathrm{u}}\frac{\alpha(\mathcal{E})d\mathcal{E}}{\mathcal{E}^2 - E^2}.
\end{equation}
Here, $c$ is the speed of light and $q$ is the magnitude of the electronic charge. The upper limit of the integral is taken to be $E_\mathrm{u} = E_\mathrm{g} + 1.5$ eV.
For the real part of the refractive index $\eta$, the dominant contribution from the high critical points $E_2 = 3$ and $E_3 = 5$ eV is captured using  \cite{RN115}
\begin{equation}
\eta(E) = \sqrt{1 + \frac{A_{\mathrm{K}2}}{E_2^2 - E^2} + \frac{A_{\mathrm{K}3}}{E_3^2 - E^2} + \frac{A_4}{E_4^2 - E^2}}.
\end{equation}
The coefficients $A_{\mathrm{K}i}$ represent the oscillator strengths (in $\mathrm{eV^2}$) related to the critical points and are quadratic functions of the temperature $T$. The last term approximates the contribution of the reststrahl band with $A_4 = 0.0002382$ ($\mathrm{eV^2}$) and $E_4 = E_\mathrm{ph} = 0.0336$ eV the optical phonon energy. All these parameters for the real part of the refractive index are determined by fitting to the precise refractive index data obtained from experiments. The dielectric permittivity for GaAs can thus be calculated using equation \eqref{epsilonNK} in a manner similar to Si.

\subsection*{Mie scattering}
The size and temperature dependent efficiencies for scattering ($Q_{\mathrm{sca}}$) and absorption ($Q_{\mathrm{abs}}$) are calculated in the temperature range $200$-$650$ K based on Mie theory using an algorithm by Wiscombe \cite{RN84}. The expressions for the efficiencies are \cite{RN84, RN102, RN135}
\begin{equation}
Q_{\mathrm{sca}} = \frac{2}{x^2} \sum^N_{n=1} (2n+1)(|a_n|^2+|b_n|^2);
\end{equation}
\begin{equation}
Q_{\mathrm{abs}} = \frac{2}{x^2} \sum^N_{n=1} (2n+1)(\mathrm{Re}(a_n+b_n) - (|a_n|^2+|b_n|^2)).
\end{equation}
The variables $a_n$ and $b_n$ represent the electric and magnetic Mie coefficients given by 
\begin{equation}
a_n  = \frac{\mu_m n_{\mathrm{r}}^2 j_n(n_{\mathrm{r}} x)[xj_n(x)]' -\mu_{\mathrm{p}} j_n(x)[n_{\mathrm{r}} xj_n(n_{\mathrm{r}} x)]'}{\mu_m n_{\mathrm{r}}^2 j_n(n_{\mathrm{r}} x)[x h_n(x)]' - \mu_{\mathrm{p}} h_n(x)[n_{\mathrm{r}} x j_n(n_{\mathrm{r}} x)]'};
\label{e_An}
\end{equation}
\begin{equation}
b_n = \frac{\mu_{\mathrm{p}} j_n(n_{\mathrm{r}} x)[x j_n(x)]' - \mu_m j_n(x)[n_{\mathrm{r}} x j_n(n_{\mathrm{r}} x)]'}{\mu_{\mathrm{p}} j_n(n_{\mathrm{r}} x)[x h_n(x)]' - \mu_m h_n(x)[n_{\mathrm{r}} x j_n(n_{\mathrm{r}} x)]'}.
\label{m_Bn}
\end{equation}
Here, $h_n(x)=j_n(x)+iy_n(x)$ are the Hankel functions of order $n$ wherein $j_n$ and $y_n$ are the spherical Bessel functions of the first and second kind respectively; $n_{\mathrm{r}} = n_{\mathrm{p}}/n_{\mathrm{m}}$ is the relative refractive index; $n_{\mathrm{p}}$ is the complex refractive index of the particle; $\mu_{\mathrm{p}}$ and $\mu_{\mathrm{m}}$ are the permeabilities of the spherical particle and the host medium respectively; and, the primes indicate differentiation of the argument in the square parentheses with respect to the size parameter $x$ ($= 2\pi r\ n_{\mathrm{m}}/\lambda$) of the particle. For our simulations, however, we assume both the host medium and the nanoparticles to be non-magnetic \textit{i.e.} $\mu_\mathrm{m} = \mu_{\mathrm{p}} = 1$. The surrounding medium is also assumed to be isotropic and non-absorbing with a constant refractive index of $n_\mathrm{m} = 1.5$. For more details on the theory and computation of Mie efficiencies the reader is referred to our previous work \cite{RN135} and the relevant text \cite{RN84, RN102}. In all our computations here, we further assume that the modeled particles are in thermal equilibrium at a given temperature and no thermal transients occur. However, the radii $r(T)$ of the Au, Si and GaAs nanoparticles are adjusted for thermal expansion in our calculations for the Mie efficiencies using
\begin{equation}
r(T) = r_\mathrm{o}(1 + \alpha_\mathcal{L}T).
\end{equation}
The coefficient of linear thermal expansion $\alpha_{\mathcal{L}}^{\mathrm{Au}}$ ($= 1.42 \times 10^{-5}$ m/K) for Au nanoparticles is assumed to be constant with temperature \cite{RN147}. The linear coefficient of thermal expansion for the Si nanoparticles is computed using \cite{RN139}
\begin{equation}
\alpha_{\mathcal{L}}^{\mathrm{Si}} = [3.725[1 - \exp(-5.88\times10^{-3}(T - 124))] + 5.548\times10^{-4}T] 10^{-6}
\end{equation}
while for the GaAs nanoparticles the thermal expansion data from Glazov \textit{et al.} \cite{RN140} is used after suitable interpolation and linear extrapolation to temperatures below 390 K. The effect of the surface scattering is taken into account for the Au nanoparticles by including a corresponding damping term in the computation of the Drude contribution to the dielectric permittivity of Au in equations \eqref{epsilonDCP} and \eqref{epsilonDL}
\begin{equation} 
\Gamma_{\mathrm{D}} \rightarrow \Gamma_{\mathrm{eff}} = \Gamma_{\mathrm{D}} + 2\pi A\frac{\hslash v_F}{r}, 
\label{g_s}
\end{equation}
where $v_\mathrm{F} = 1.4\times10^6$ m/s and $A = 0.33$ is a geometric form factor estimated using the best fitting value for absorption spectra from individual gold nanoparticles \cite{RN147}. However, this term is neglected in our calculations for the intrinsic Si and GaAs nanoparticles because of the characteristic low density of charge carriers.

\section*{RESULTS AND DISCUSSION}
\subsection*{Complex refractive indices}
\begin{figure}[ht!]	
\centering
\includegraphics[scale=0.72]{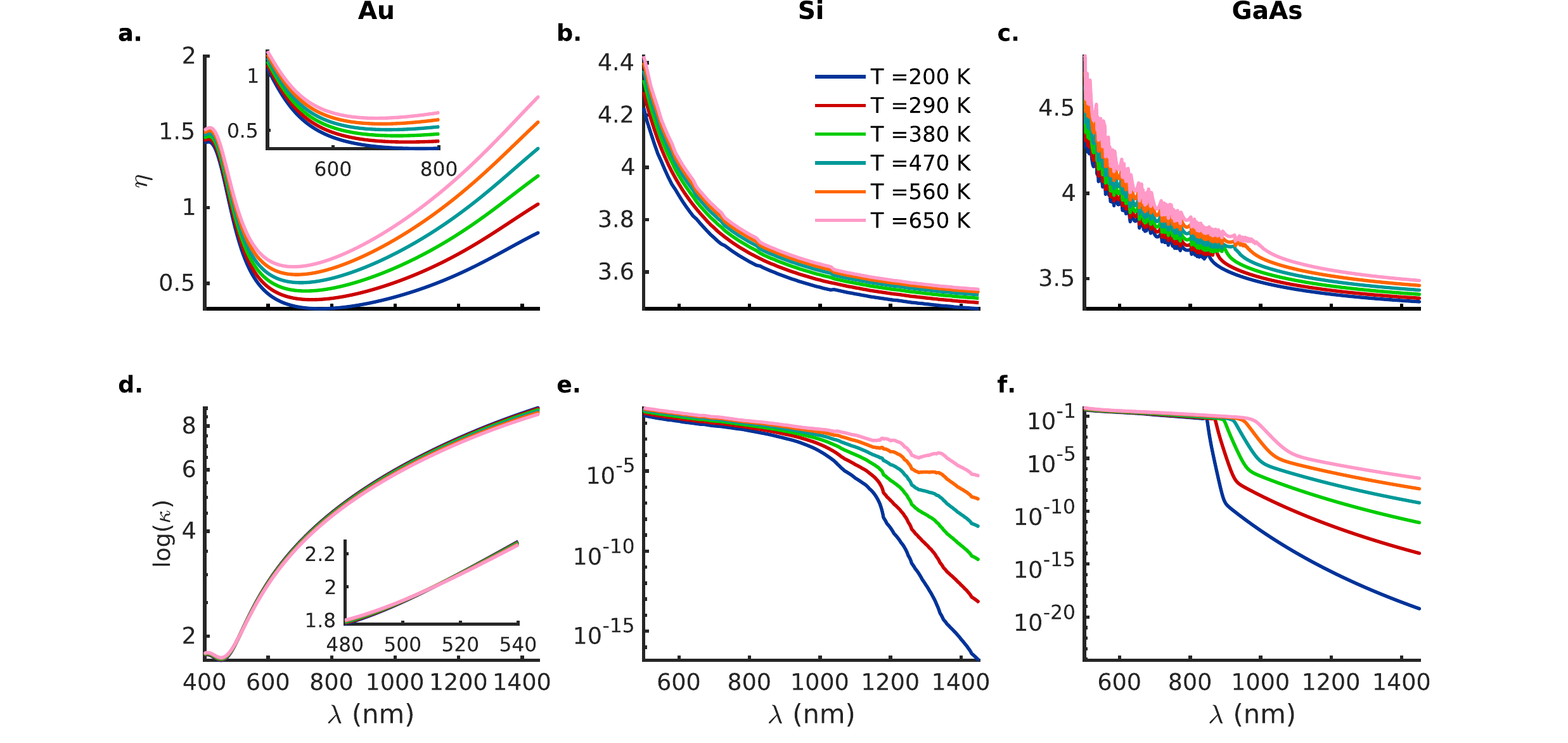} 
\caption{Comparison of the temperature dependent (a-c) real ($\eta$) and (d-f) imaginary ($\kappa$) parts of the refractive indices for Au (DL model), Si and GaAs respectively. In (a-c), $\eta$ is seen to increase monotonically with temperature, $T$, for all materials. The inset in (d) shows the non-monotonic behavior of $\kappa$ for Au: $\kappa$ increases with temperature at shorter wavelengths followed by a decrease at the longer wavelengths. This is in contrast with the monotonic increase of $\kappa$ with temperature for Si and GaAs in (e-f).}
\label{compRefInd} 
\end{figure}

We first examine and compare the results for the temperature dependent complex refractive indices for the materials Au, Si and GaAs. Figure \ref{compRefInd}a-c, d-f shows the real ($\eta$) and the imaginary ($\kappa$) parts of the refractive indices for Au, Si and GaAs respectively. The results in Figure \ref{compRefInd}a,d for $\eta$ and $\kappa$ have been computed using the DL model of temperature dependent dielectric permittivity for Au while Figure S1a,b shows the results from the DCP model. For a temperature $T$, Figure \ref{compRefInd}a shows that $\eta$ for Au initially decreases in the interval $\lambda = 400$-$760$ nm (see inset, Figure \ref{compRefInd}a) and then increases continuously up to $1450$ nm with an increase in the wavelength of the incident radiation. A similar behavior is seen for $\kappa$ wherein for a given temperature $T$, it decreases with an increase in wavelength until $450$ nm and monotonically increases thereafter. However, $\eta$ as a function of the temperature increases monotonically with an increase in the temperature $T$ from $200$ to $650$ K at any given wavelength in the $400$-$1450$ nm range. In contrast, $\kappa$ increases for wavelengths in the range $400$-$510$ nm and then decreases until $\lambda = 1450$ nm with an increase in the temperature from $200$ to $650$ K (see inset, Figure \ref{compRefInd}b). 

The results for the refractive index from the DCP model (Supplementary Information, Figure S1a,b ) show a variation that is very similar to the DL model as a function of the wavelength. Unlike the DL model, however, the results from the DCP model show a highly non-monotonic variation in $\eta$ and $\kappa$ with an increase in temperature from $200$-$650$ K (Figure S1a,b). In the wavelength interval $400$-$500$ nm, $\eta$ decreases with temperature until 560 K and thereafter increases slightly between $560$ and $650$ K. Then there occurs a transition region in the interval between $\lambda = 500$-$740$ nm (see inset: Figure S1a) wherein $\eta$ initially increases with an increase in $T$ from $200$ to $470$ K, then decreases between $470$-$560$ K followed by an increase again for $T > 560$ K. Between $\lambda = 740$-$1450$ nm, $\eta$ increases monotonically with an increase in temperature. Also, notably the relative increase in $\eta$ is greater at longer wavelengths (beyond $740$ nm) with nearly a doubling of the $\eta$ value between $200$ and $650$ K at $\lambda = 1450$ nm. As opposed to the decrease of $\kappa$ with an increase in the temperature in the DL model, the $\kappa$ for Au in the DCP model increases monotonically at all wavelengths (except for a small region between $400$-$425$ nm) for temperatures between $200$ and $560$ K followed by a saturation between $560$ and $650$ K (inset: Figure S1b). 

This temperature dependent behavior of the refractive index, at longer wavelengths and away from the interband transitions at $\sim$$470$ and $\sim$$330$ nm in the DCP model, can be attributed to the changes in the plasma frequency $\omega_\mathrm{p}$ and the Drude damping $\Gamma_\mathrm{D}$ [$=\hslash/\tau_\mathrm{D}$, $\tau_\mathrm{D}$ $\equiv$ Drude relaxation time]. Reddy \textit{et al.} \cite{RN114} demonstrate through their \textit{experimental} work that the temperature dependence of $\omega_\mathrm{p}$ and $\Gamma_\mathrm{D}$ is a direct consequence of (a) the decrease in carrier density $N$ [$\propto(1 + 3\alpha_\mathcal{L}\Delta T)^{-1}$] due to thermal expansion, (b) decrease in the effective mass $m^*$, and, (c) an increase in the electron-phonon (\textit{e-$\phi$}) interaction with an increase in temperature. The counteracting effect of the decrease in $N$ and $m^*$ is reflected in the initial increase of $\omega_\mathrm{p}$ with rising temperatures up to $\sim$$473$ K wherein the decrease in $m^*$ is dominant. At still higher temperatures ($T > 473$) K, a decrease in $\omega_\mathrm{p}$ follows due to a decrease in the charge carrier density as a result of thermal expansion \cite{RN114}. In contrast, the temperature dependence of the plasma frequency in the DL model has contributions only from the changes in the carrier concentration $N$ as a result of thermal expansion while the effective mass of the free electrons is assumed to be independent of the temperature \cite{RN147, RN118}. In the DL model, the temperature dependence of the \textit{e-e} interactions shown in equation \eqref{g_ee} is quadratic in $T$ but is masked by the high frequencies ($\omega$) of the incident radiation ($\lambda = 400$-$1450$ nm) \cite{RN147, RN132, RN119}. Reddy \textit{et al.} show that in the DCP model the contribution of the \textit{e-$\phi$} interaction to $\Gamma_\mathrm{D}$ for temperatures $T$ above the Debye temperature ($\theta = 170$ K) is linear and directly proportional to $T$ thereby resulting in a steady increase in $\Gamma_\mathrm{D}$ with a rise in $T$ \cite{RN114}. In the DL model, however, no such assumption is made and the temperature dependence of the damping factor $\Gamma_\mathrm{e\phi}$ in equation \eqref{g_eph} is computed without any assumption of linearity \cite{RN147, RN132, RN119}.

\begin{figure}[ht!]	
\centering
\includegraphics[scale=0.64]{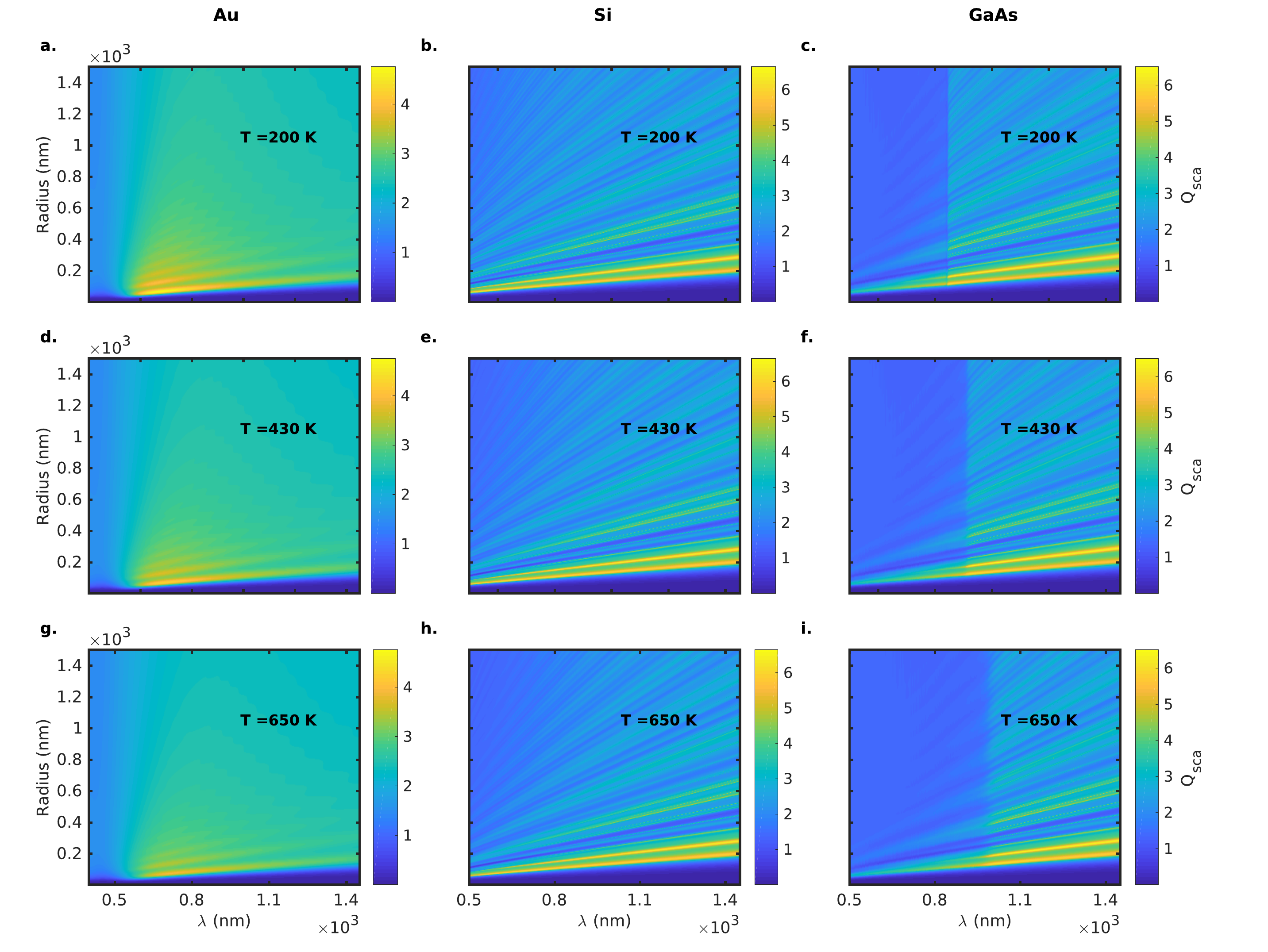} 
\caption{Mie scattering efficiency, $Q_\mathrm{sca}$, as a function of the wavelength, $\lambda$, of the incident radiation and the particle radii $r$ at three different temperatures: $T=200$ (a-c), $430$ (d-f) and $650$ K (g-i) for Au (DL model), Si and GaAs particles, respectively. The strength of the resonances in $Q_\mathrm{sca}$ for Au particles shows a decline with an increase in the temperature, $T$, whereas the Si and GaAs particles show no significant deterioration. (See Figure \ref{mQsca} also.)}
\label{qsca2d} 
\end{figure}

Considering next the semiconductors Si and GaAs, Figures \ref{compRefInd}b,e and \ref{compRefInd}c,f respectively show that the increase in $\eta$ and $\kappa$ with increasing temperatures is monotonic across the entire wavelength range of $500$-$1450$ nm. The extinction coefficient for Si ($\kappa\sim$$0.1$-$0.001$) and GaAs ($\kappa\sim$$0.6$-$0.1$) is also observed to stay almost constant with temperature for the shorter wavelengths within the absorption band. However, at the longest of wavelengths ($\lambda$$\sim$$1450$ nm), $\kappa$ continually increases with temperature from low values on the order of $10^{-17}$ to about $10^{-6}$ and $10^{-20}$ to $10^{-7}$ for Si and GaAs, respectively. Thus, with the values of $\kappa$ for Au lying between $6$-$10$ at the longer wavelengths, the $\kappa$-values for Si and GaAs are still about $7$-$8$ orders of magnitude smaller compared to those for Au at the highest temperature ($T = 650$ K). Such low values of $\kappa$ for the \textit{intrinsic} Si and GaAs point to the low number of charge carriers generated for longer wavelengths of the incident radiation away from the absorption band. At the same time, an almost $13$-$14$ orders of magnitude increase in $\kappa$ points to the enhanced \textit{e-$\phi$} coupling at elevated temperatures. This leads to an increase in the intrinsic carrier concentrations $\rho_i$ [$\propto \exp(-E_\mathrm{g}(T)/2k_\mathrm{B}T)$] due to bandgap shrinkage and elevated temperatures. Also, between Si and GaAs, the sharp band-edge representing a steep increase in $\kappa$ is clearly visible for the direct bandgap GaAs. This bandgap shrinks with an increase in the temperature as evidenced by the significant red-shift of the absorption band-edge from $\lambda \approx 900$ nm at $200$ K to $\lambda \approx 1090$ nm at $650$ K. In contrast, there exists no noticeable sharp increase in $\kappa$ for the indirect bandgap Si and instead a gradual increase representing phonon-mediated interband transitions is observed. Although, similar to GaAs, an increase in temperature is accompanied by a shrinking of the bandgap in Si as well.

\begin{figure}[ht!]	
\centering
\includegraphics[scale=0.64]{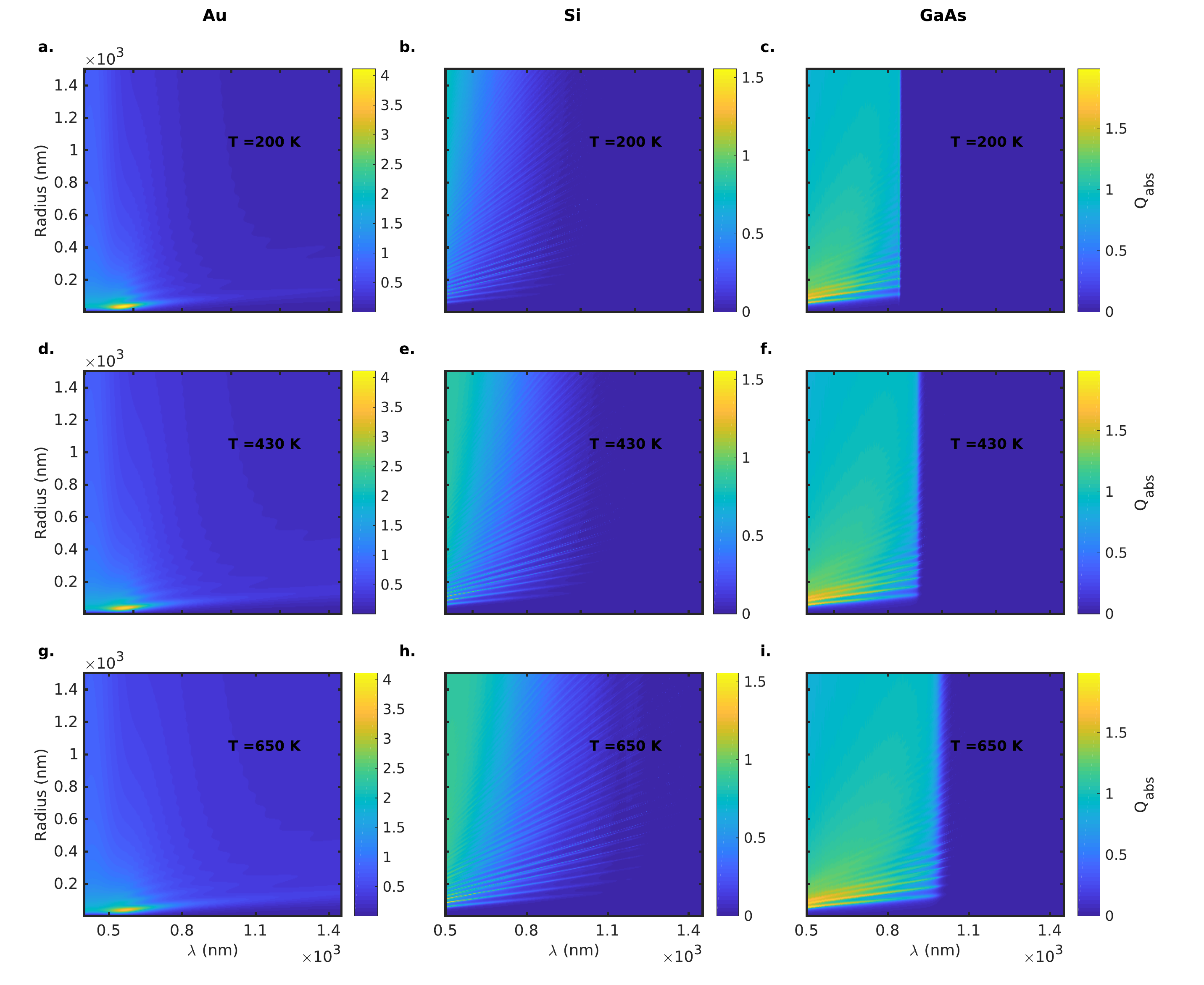} 
\caption{Mie absorption efficiency, $Q_\mathrm{abs}$, as a function of the wavelength, $\lambda$, of the incident radiation and the particle radii $r$ at three different temperatures $T=200$ (a-c), $430$ (d-f) and $650$ K (g-i) for Au (DL model), Si and GaAs particles, respectively. A decrease in $Q_\mathrm{abs}$ for the smaller Au particles is observed with a rise in temperature at shorter wavelengths in contrast to an increase at the longer wavelengths. The Si and GaAs particles exhibit no significant decline in $Q_\mathrm{abs}$ at elevated temperatures and a clear redshift of the absorption resonances is seen. (See Figure \ref{mQabs} also.)}
\label{qabs2d} 
\end{figure}

\subsection*{Mie scattering and absorption efficiencies}
Figures \ref{qsca2d} and \ref{qabs2d} show the variations in the Mie scattering ($Q_\mathrm{sca}$) and absorption ($Q_\mathrm{abs}$) efficiencies, respectively, at three different temperatures $T=200$, $430$ and $650$ K as a function of the wavelength, $\lambda$, of the incident radiation and the particle radii in the $5$-$1500$ nm range for Au (DL model; see Figures S2 and S3 for the DCP model), Si and GaAs particles. Here, two types of broadening and a shift of spectral features are evident - one arising from the increase in the particle radii and the other due to an increase in the temperature. The second temperature dependent contribution to the broadening and shifting of the spectral features at elevated temperatures is straight-forward to understand and interpret for the Au (DL model), Si and GaAs particles. However, this is rendered rather complicated in the case of the Au (DCP model) nanoparticles because of the non-monotonic behavior of the dielectric permittivity ($\epsilon$) or refractive index ($n_\mathrm{p}$) with rising temperatures in the DCP model (Figures S1 and S5). Therefore, we first address the effects arising due to an increase in the particle radii and leave a discussion on the temperature effects for later when we consider the thermoplasmonic response of individual nanoparticles with a given radii. 

A broadening and redshifting of the spectral features due to increasing particle radii are observed for Au, Si and GaAs particles alike as a result of the weakening of the restoring force that drives the plasmonic resonances (Figures \ref{qsca2d}, \ref{qabs2d} and S2-S3). The increase in the particle radii leads to an increased distance between the oscillating charges on the opposite sides of the particle and therefore lower plasmon energies corresponding to a redshift. This particle size-dependent redshift can be better understood when one considers the small particle limit such that the size parameter $x \ll 1$. The resonance condition for the $n^\mathrm{th}$ order electric modes $a_n$ in the case of a vanishingly small particle ($x \rightarrow 0$), wherein the magnetic modes can be neglected, requires that the denominator in equation \eqref{e_An} goes to zero and is given by \cite{RN102, RN135}
\begin{equation}
\epsilon(\omega_n) = -\frac{n+1}{n}\epsilon_\mathrm{m}.
\label{resN}
\end{equation}
Here, $\epsilon_\mathrm{m} = 2.25$ is the dielectric permittivity for the non-absorbing host medium. The frequencies ($\omega_n$) obtained from the above resonance condition are \textit{complex} and therefore the associated modes are referred to as \textit{virtual} \cite{RN102, RN135}. Thus, for real materials this condition can only be satisfied approximately. In such cases, the real frequency that approximates this resonance condition closest is referred to as the Fr$\mathrm{\ddot{o}}$hlich frequency \cite{RN102}. Furthermore, the imaginary part $\epsilon''$ of the dielectric permittivity is required to vanish here because the left-hand side of equation \eqref{resN} is complex while the right-hand side is real. 

\begin{figure}[ht!]	
\centering
\includegraphics[scale=0.64]{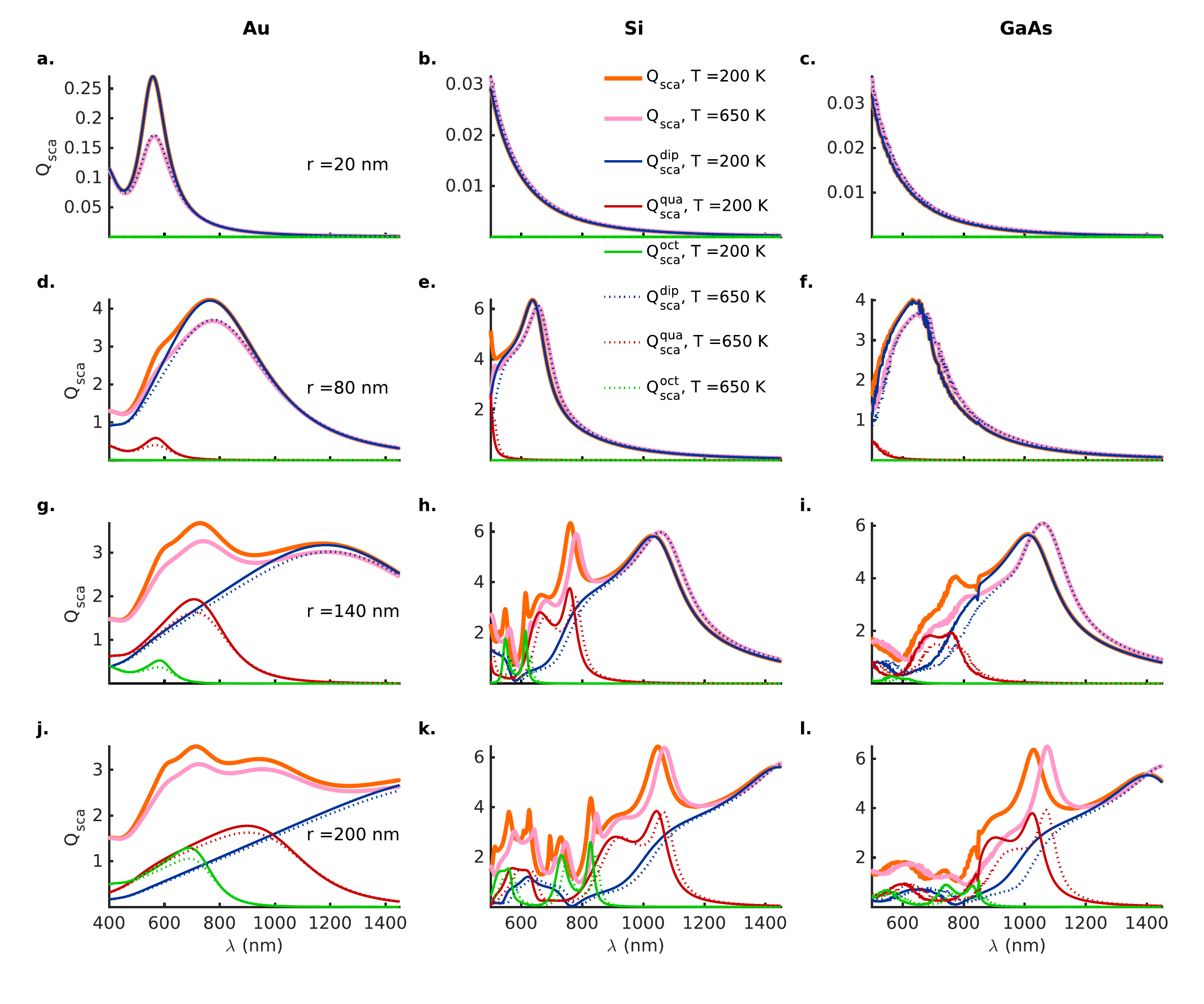} 
\caption{Mie scattering efficiency ($Q_\mathrm{sca}$) with dipole (blue), quadrupole (red) and octupole (green) contributions as a function of the wavelength, $\lambda$, for (a, d, g, j) Au (DL model), (b, e, h, k) Si and (c, f, i, l) GaAs nanoparticles of radii $r = 20, 80, 140$ and $200$ nm, respectively, at temperatures $T = 200$ (solid lines) and $650$ (dotted lines) K. The thick orange and pink solid lines represent the total scattering efficiencies, $Q_\mathrm{sca}$, at temperatures $T = 200$ and $650$ K, respectively. The sharp higher order resonances in Si and GaAs nanoparticles against a background of the broad dipole modes give rise to Fano resonances that do not decay significantly with an increase in the temperature. In contrast, the Mie resonances in the Au nanoparticles exhibit massive Drude broadening and a deterioration in strength with temperature increase due to dissipative damping. Here, the Mie computations for the nanoparticles of different sizes take into account their thermal expansion, although the text labels indicate the values for nanoparticle radii at $200$ K.}
\label{mQsca} 
\end{figure}

For a finite but small size parameter ($x \ll 1$) (the case of interest here), the resonance condition for the Fr$\mathrm{\ddot{o}}$hlich mode is obtained by a power series expansion of the Hankel ($h_n$) and Bessel ($j_n$) functions in the denominator of equation \eqref{e_An} to second order in the size parameter $x$. The equation \eqref{resN} then gets modified to \cite{RN102, RN135, RN149}
\begin{equation}
\epsilon' = -(2 + 12x^2/5)\epsilon_{\mathrm{m}}.
\label{frohlich}
\end{equation}    
Now, an examination of the dielectric permittivity in Figures S4a-c and S5a reveals that for all three materials, metallic Au (DL and DCP models) and the semiconducting Si and GaAs, the real part $\epsilon'$ [$=\eta^2 - \kappa^2$] of the dielectric permittivity increases with decreasing wavelength in the range $\lambda = 400$-$1450$ nm and $500$-$1450$ nm, respectively. Therefore, since the right-hand sides of equations \eqref{resN} and \eqref{frohlich} are negative, an increase in the size parameter $x$ due to an increase in the particle radii redshifts the resonance frequency for the electric modes $a_n$. 

A further comparison of $Q_\mathrm{sca}$ in Figure \ref{qsca2d} for the Au nanoparticles (DL model) with the Si and GaAs nanoparticles shows a general decline in the scattering efficiency for the Au nanoparticles with an increase in the temperature as opposed to steady values of $Q_\mathrm{sca}$ for the Si and GaAs nanoparticles. Also, it is observed that the results from the DCP model (Figure S2) show a stronger scattering efficiency than the $Q_\mathrm{sca}$ for the DL model in Figure \ref{qsca2d}. Next, the distinct variations in the Mie absortpion efficiency $Q_\mathrm{abs}$ are shown in Figure \ref{qabs2d} as a function of the wavelength $\lambda$ and the particle radii $r$ varying between $5$-$1500$ nm at three different temperatures ($T=200$, $430$ and $650$ K) for Au (DL model), Si and GaAs particles. In Figure \ref{qabs2d}a,d,g, the Au nanoparticles exhibit increased absorption efficiencies at longer wavelengths with an increase in temperature for the larger particles while for the smaller Au nanoparticles there occurs a decrease in $Q_\mathrm{abs}$. Figure \ref{qabs2d}b,e,h shows that for the Si particles, the absorption appears diffuse characterized by a redshift to longer wavelengths with an increase in temperature. The diffuseness and the redshift of the $Q_\mathrm{abs}$ spectra are a result of the phonon-mediated interband transitions due to an increase in the \textit{e-$\phi$} interaction and a shrinking of the indirect bandgap at high temperatures, respectively. With the notable exception of a sharp absorption band-edge, a similar spectral redshift is also observed in $Q_\mathrm{abs}$ for the GaAs particles due to bandgap shrinkage (Figure \ref{qabs2d}c,f,i). The sharp band-edge is a consequence of the direct interband transitions in GaAs and becomes progressively diffuse on account of increased \textit{e-$\phi$} coupling at elevated temperatures (compare subplots, Figure \ref{qabs2d}c,f,i). 

\begin{figure}[ht!]	
\centering
\includegraphics[scale=0.64]{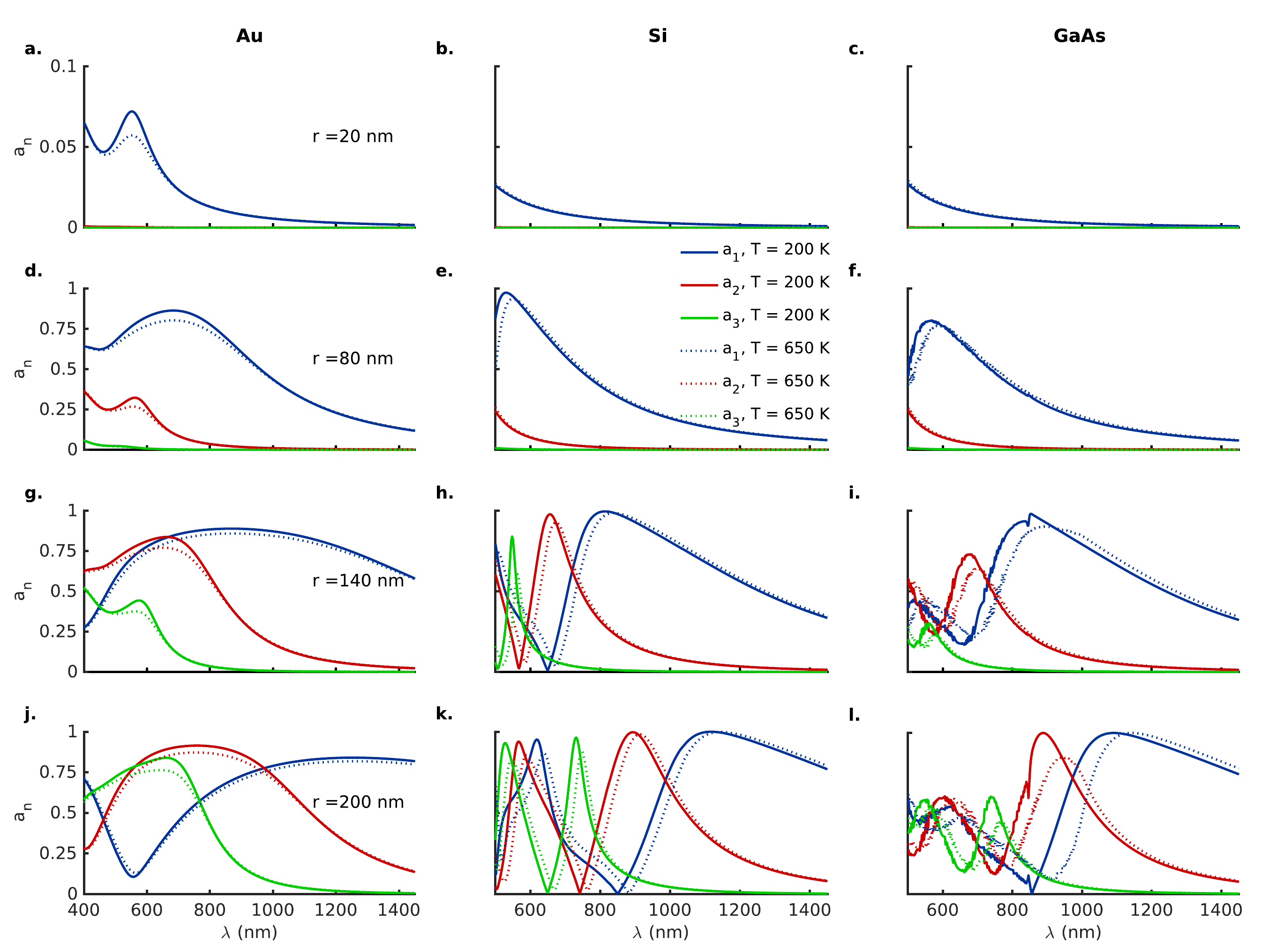} 
\caption{Electric dipole (blue), quadrupole (red) and octupole (green) Mie modes ($a_n$) as a function of the wavelength, $\lambda$, for (a, d, g, j) Au (DL model), (b, e, h, k) Si and (c, f, i, l) GaAs nanoparticles of radii $r = 20, 80, 140$ and $200$ nm, respectively, at temperatures $T = 200$ (solid lines) and $650$ (dotted lines) K. A broadening and redshifting of the electric Mie resonances is observed with increasing particle size and temperature for all materials. The larger Si and GaAs nanoparticles exhibit sharp higher order electric resonances. These, alongwith the magnetic Mie modes ($b_n$) (Figure \ref{b_n}), are responsible for giving rise to Fano resonances in $Q_\mathrm{sca}$ and $Q_\mathrm{abs}$ for Si and GaAs nanoparticles (Figures \ref{mQsca} and \ref{mQabs}). Here, the Mie computations for the nanoparticles of different sizes take into account their thermal expansion, although the text labels indicate the values for nanoparticle radii at $200$ K.}
\label{a_n} 
\end{figure}

\begin{figure}[ht!]	
\centering
\includegraphics[scale=0.64]{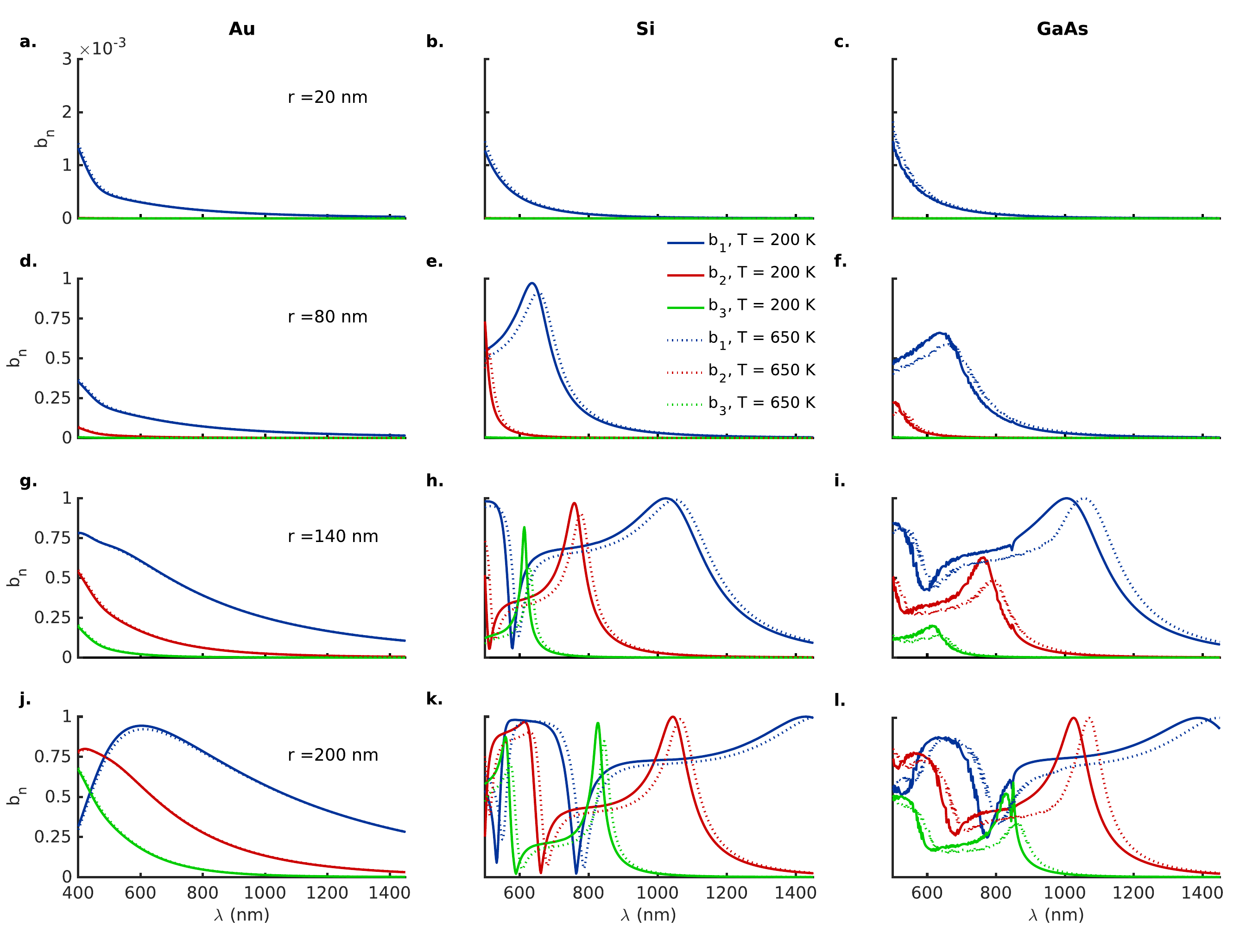} 
\caption{Magnetic dipole (blue), quadrupole (red) and octupole (green) Mie modes ($b_n$) as a function of the wavelength, $\lambda$, for (a, d, g, j) Au (DL model), (b, e, h, k) Si and (c, f, i, l) GaAs nanoparticles of radii $r = 20, 80, 140$ and $200$ nm, respectively, at temperatures $T = 200$ (solid lines) and $650$ (dotted lines) K. Similar to the electric Mie resonances $a_n$ (Figure \ref{a_n}), the magnetic resonances also broaden and redshift with an increase in particle size and temperature for all materials. The larger Si and GaAs nanoparticles exhibit sharp higher order magnetic Mie resonances. These, alongwith the electric Mie modes ($a_n$) (Figure \ref{a_n}), are responsible for giving rise to the Fano resonances in $Q_\mathrm{sca}$ and $Q_\mathrm{abs}$ for Si and GaAs nanoparticles (Figures \ref{mQsca} and \ref{mQabs}). Here, the Mie computations for the nanoparticles of different sizes take into account their thermal expansion, although the text labels indicate the values for nanoparticle radii at $200$ K.}
\label{b_n} 
\end{figure}

We next examine the contributions of the different Mie modes to $Q_\mathrm{sca}$ to further understand the effect of the temperature and particle size on its spectral behavior. Figure \ref{mQsca} presents the contributions from the dipole, quadrupole and octupole modes (order $n \le 3$) to the total scattering efficiencies for particles of radii $20, 80, 140$ and $200$ nm at $T = 200$ and $650$ K. Note, that for temperatures greater than $200$ K we take into account the thermal expansion of the nanoparticles. Figure \ref{mQsca}a-c and a comparison of Figures \ref{a_n}a-c and \ref{b_n}a-c show that for the smaller $20$ nm particles the strongest and the most significant contribution to $Q_\mathrm{sca}$ comes from the electric dipole mode for all three types of nanoparticles (Au (DL model), Si and GaAs). Upon increasing the particle radii to $80$ nm (Figure \ref{mQsca}d-f and Figures \ref{a_n}d-f, \ref{b_n}d-f), the dipole contribution to $Q_\mathrm{sca}$ begins to exhibit a strong and distinct resonance that broadens and redshifts with an increase in temperature from $200$ to $650$ K. Additionally, there appears a weak quadrupole contribution from both magnetic and electric modes at short wavelengths that is strongest for the Si nanoparticles (Figures \ref{mQsca}e, \ref{a_n}e and \ref{b_n}e).  For the $80$ nm Si and GaAs particles a large contribution to $Q_\mathrm{sca}$ comes from the magnetic modes that exhibit sharper resonances and occur at longer wavelengths than the electric dipole modes (Figures \ref{a_n}d-f and \ref{b_n}d-f). In all cases, the magnetic Mie resonances of the same order as the electric modes occur at longer wavelengths and hence have lower associated energies. However, in the case of the $80$ nm Au nanoparticles the strongest contributor to the resonance in $Q_\mathrm{sca}$ is by far still the electric dipole mode (Figures \ref{mQsca}d, \ref{a_n}d and \ref{b_n}d). With an increase in the temperature there is also a weakening of the dipole contribution in the 80 nm nanoparticles for all materials. 

As one transitions to the larger $140$ and $200$ nm particles, a strengthening of the quadrupole and octupole modes is observed in all materials (Figures \ref{mQsca}g-l, \ref{a_n}g-l and \ref{b_n}g-l). In Au nanoparticles, the quadrupole and octupole modes are also very broad and contribute to the already massive Drude broadening from the dipole modes (Figures \ref{mQsca}g,j, \ref{a_n}g,j and \ref{b_n}g,j). The largest of the Au nanoparticles with radii equal to $140$ and $200$ nm have significant contributions to $Q_\mathrm{sca}$ from the magnetic modes $b_n$. However, these are still weaker than contributions from the electric modes $a_n$ (Figures \ref{mQsca}g,j, \ref{a_n}g,j and \ref{b_n}g,j). In contrast, sharp magnetic resonances $b_n$ occur against a background of the broad electric modes $a_n$ in the larger ($r = 140$, $200$ nm) Si and GaAs nanoparticles. This gives rise to Fano resonances in the corresponding spectrum for $Q_\mathrm{sca}$ (Figures \ref{mQsca}h-i,k-l, \ref{a_n}h-i,k-l  and \ref{b_n}h-i,k-l). The Fano resonances from the octupole magnetic modes ($b_3$), in particular, are observed to be especially sharp for the Si nanoparticles compared to those for the GaAs nanoparticles (Figures \ref{mQsca}h-i,k-l, \ref{a_n} and \ref{b_n}). For all radii and materials a redshifting, weakening and broadening of the Mie resonances of all orders is observed with an increase in temperature from $200$ to $650$ K. This is consistent with the increase in the Drude damping associated with free carrier absorption for Au and the increased \textit{e-e} and \textit{e-$\phi$} interactions in the semiconductors at elevated temperatures \cite{RN114,RN147}. 

\begin{figure}[ht!]	
\centering
\includegraphics[scale=0.64]{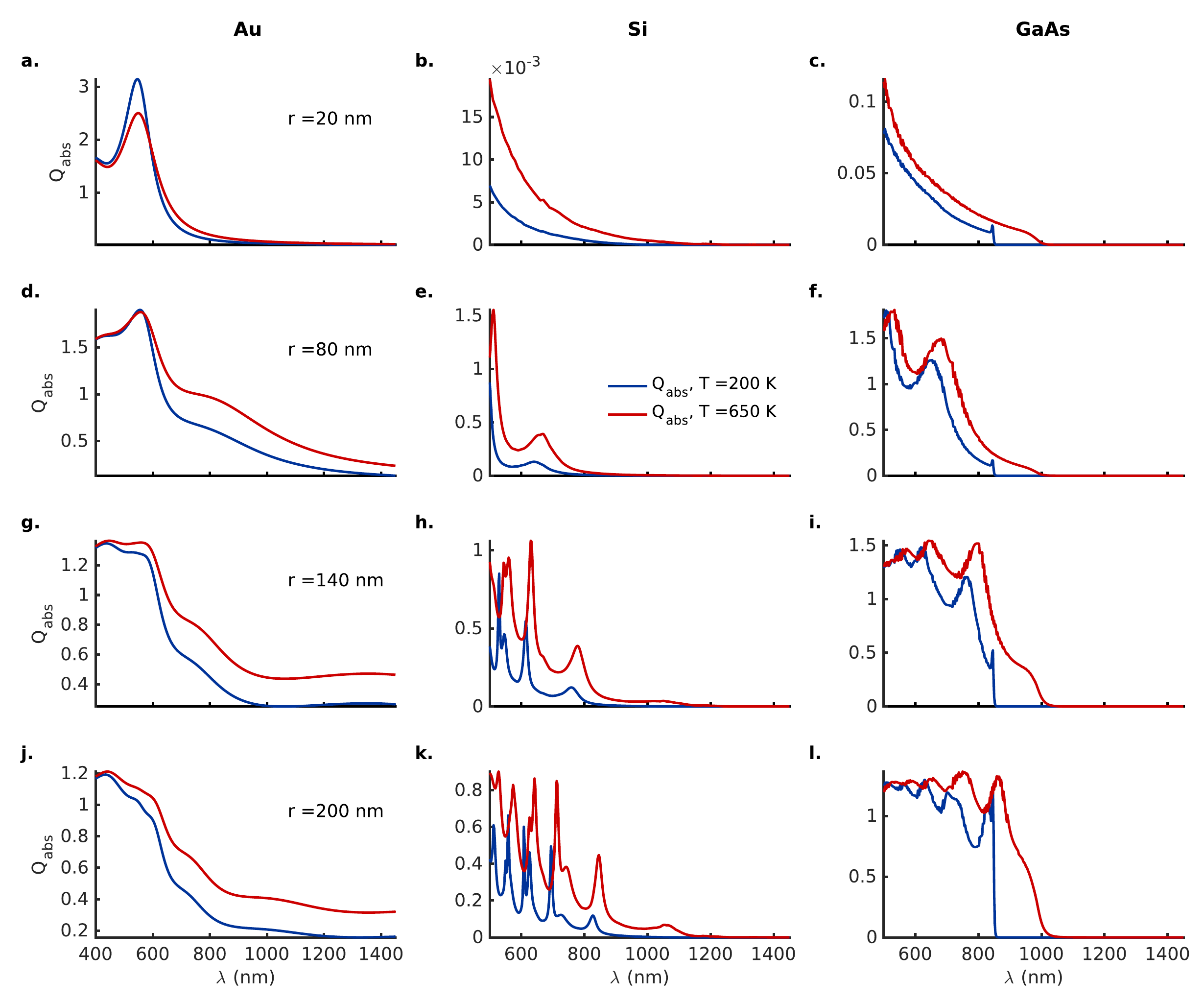} 
\caption{Mie absorption efficiency ($Q_\mathrm{abs}$) as a function of the wavelength $\lambda$ for (a, d, g, j) Au (DL model), (b, e, h, k) Si and (c, f, i, l) GaAs nanoparticles of radii $r = 20, 80, 140$ and $200$ nm, respectively, at temperatures $T = 200$ (blue) and $650$ (red) K. For the Au nanoparticles, the highest $Q_\mathrm{abs}$ is observed for the smallest 20 nm particles. An increase in $Q_\mathrm{abs}$ with temperature for the larger Au nanoparticles is evident at the longer wavelengths due to uniform Drude broadening. The Si and GaAs nanoparticles exhibit higher $Q_\mathrm{abs}$ for the larger particles at short wavelengths and negligible values at longer wavelengths with increase in temperature. A redshift in the resonances is also observed at elevated temperatures for the semiconductor nanoparticles. Here, the Mie computations for the nanoparticles of different sizes take into account their thermal expansion, although the text labels indicate the values for nanoparticle radii at $200$ K.}
\label{mQabs} 
\end{figure}

Figure \ref{mQabs} shows the absorption efficiency $Q_\mathrm{abs}$ as a function of $\lambda$ for the Au (DL model), Si and GaAs nanoparticles of radii $r = 20, 80, 140$ and $200$ nm at the temperatures $200$ and $650$ K. Similar to the results for $Q_\mathrm{sca}$ (Figure \ref{mQsca}a-c), the strongest contribution to $Q_\mathrm{abs}$ for the smaller $20$ nm particles comes from the electric dipole modes $a_1$ for the nanoparticles of all three materials (Figure \ref{a_n}a-c). However, a clear difference in $Q_\mathrm{abs}$ between the metallic Au and the semiconductor nanoparticles is seen with an increase in the temperature from $200$ to $650$K. The plasmonic resonance peak in $Q_\mathrm{abs}$ for the $20$ nm Au nanoparticles broadens and decreases in height with an increase in the temperature. In contrast, the absorption efficiency $Q_\mathrm{abs}$ for the $20$ nm Si and GaAs particles exhibits an increase at higher temperatures. We note here that Yeshchenko and coworkers have also reported similar experimental results for the temperature dependence of plasmonic resonances in Au nanoparticles ($r = 10$ nm) that are embedded in a silica matrix \cite{RN122}. Here, the decrease in $Q_\mathrm{abs}$ for the $20$ nm Au nanoparticles occurs because of an increase in the Drude damping from the greater \textit{e-e} and \textit{e-$\phi$} coupling at elevated temperatures (equations \eqref{g_ee} and \eqref{g_eph}). On the other hand, the increase in the $Q_\mathrm{abs}$ for the Si and GaAs nanoparticles is driven by the bandgap shrinkage effect and the increased \textit{e-$\phi$} interactions associated with the rise in temperature from $200$ to $650$ K. This characteristic redshift and broadening of the absorption band at high temperatures is evident in the absorption efficiencies for the larger ($80$, $140$, and $200$ nm) Si and GaAs nanoparticles as well (Figures \ref{mQabs}e-f,h-i,k-l, \ref{a_n}e-f,h-i,k-l, and \ref{b_n}e-f,h-i,k-l, respectively). The appearance of the broad higher order electric modes ($n \ge 2$) at shorter wavelengths in the $140$ and $200$ nm Au particles against a background of the mostly featureless and broad magnetic modes (except for the magnetic dipole mode $b_1$ for the $200$ nm Au particle) leads to small humps in $Q_\mathrm{abs}$. Unlike the smaller ($r=20$ nm) Au nanoparticles, these strong contributions from the higher order electric modes to $Q_\mathrm{abs}$ in the larger Au nanoparticles also counteract the decrease in the height of the plasmonic resonance peak due to Drude damping at elevated temperatures. The stronger absorption efficiency $Q_\mathrm{abs}$ for the large Au nanoparticles ($r = 80, 140$ and $200$ nm; Figure \ref{mQabs}d,g,j) observed at the long wavelengths, away from the interband transitions, can be attributed to the accompanying free carrier absorption and the increased \textit{e-$\phi$} interaction at higher temperatures. In contrast to the broad resonance absorption observed for the Au nanoparticles due to the electric modes, the Si and GaAs nanoparticles exhibit Fano resonances (Figure\ref{mQabs}e-f,h-i,k-l). These arise due to contributions from the higher order resonant Mie modes ($a_n$ and $b_n$) that occur against a background of the broad quadrupole or dipole modes (Figures \ref{a_n}e-f,h-i,k-l and \ref{b_n}e-f,h-i,k-l). In particular, for the larger Si nanoparticles ($r = 140$ and $200$ nm), the Fano resonances are sharper and their number increases with the size of the nanoparticle (Figures \ref{a_n}h-i,k-l and \ref{b_n}h-i,k-l, respectively) \cite{RN135}. 

The electric ($a_n$) or magnetic ($b_n$) modes that contribute the most to the resonant plasmonic absorption in the nanoparticles of these three materials are the higher order ($n \ge 2$) Mie modes that occur at wavelengths close to the interband transitions (Figures \ref{a_n} and \ref{b_n}). This is not a coincidence as the maximal contribution $Q_\mathrm{abs}^{n,\mathrm{max}}$ of the $n^\mathrm{th}$ order Mie mode to $Q_\mathrm{abs}$ for a small but finite-sized particle is given by \cite{RN123}
\begin{equation}
Q_\mathrm{abs}^{n,\mathrm{max}} = \frac{1}{x^2}\bigg[n + \frac{1}{2}\bigg].
\label{qmax}
\end{equation} 
Thus, for small but finite-sized particles the maximal absorption efficiency $Q_\mathrm{abs}^{n,\mathrm{max}}$ for a Mie mode is directly proportional to its order $n$. Again, the resonance condition (equation \eqref{resN}) for the higher order Mie modes also shows that these resonances always occur at wavelengths less than those for the dipole modes. This is because for most materials, including Au, Si and GaAs considered here, the real part $\epsilon'$ of the dielectric permittivity is an increasing function of the decreasing wavelength of the incident radiation  (Figures S4a-c and S5a). The factor $(n+1)/n \in [2, 1]$ for $n \in [1, \infty]$ and the negative right hand side in equation \eqref{resN} when combined with the preceding observation ensure that the higher order Mie resonances appear blueshifted relative to the dipole modes $a_1$ and $b_1$.
 
\subsubsection*{Drude-Lorentz Vs Drude and critical points model}
The Mie resonances in the DL model for Au nanoparticles discussed above exhibit a redshift with an increase in the temperature (Figures 4-7a,d,g,j). In contrast, the results from Mie computations based on the DCP model estimated by Reddy \textit{et al.} \cite{RN114} present a blueshift (Figures S6-S9). The redshift of the resonances in $Q_\mathrm{sca}$ and $Q_\mathrm{abs}$ (Figures \ref{mQsca}a,d,g,j and \ref{mQabs}a,d,g,j respectively), obtained using the DL model here, is consistent with the experimental results reported by Yeshchenko \textit{et al.} \cite{RN122} on the temperature dependent shift of the plasmon resonance for Au nanoparticles embedded in a silica matrix. The difference in the spectral shift of the Mie resonances between the results from the DL (Figures \ref{mQsca}-\ref{mQabs}) and DCP (Figures S6-S9) models arises primarily because the real part $\epsilon'$ of the dielectric permittivity in the DCP model (Figure S5a) decreases with an increase in the temperature as opposed to an increase in the DL model (Figure S4a). The real part $\epsilon'$ of the permittivity, we recall, is an increasing function of the decreasing wavelength of the incident radiation (Figures S4a and S5a). Thus, the blueshift of the Mie resonances for the Au nanoparticles in the DCP model (Figures S6-S9) is a direct consequence of (a) the resonance condition in equation \eqref{resN} for the electric modes $a_n$, and, (b) the decrease in $\epsilon'$ as a function of the rising temperature (Figure S5a). Note here that the DCP model for the temperature dependent dielectric permittivity of Au is estimated from the spectroscopic ellipsometry measurements of 200 nm thick Au films deposited on a mica substrate under ambient conditions \cite{RN114}. Similarly, the DL model estimated by Rakic \textit{et al.} \cite{RN118} also makes use of the ellipsometric measurements of Au thin films for the estimation of the Drude and Lorentz oscillator parameters at room temperature. However, the key difference between the two models is that the Lorentz oscillator parameters for the interband transitions in the DL model \cite{RN118} are independent of temperature whereas the critical point oscillator parameters in the DCP model \cite{RN114} depend on the temperature. An additional difference between the two models comes from the temperature dependence of the Drude contribution to the permittivity. In the DL model, the Drude contribution from the \textit{e-e} and \textit{e-$\phi$} interactions (equations \eqref{g_ee} and \eqref{g_eph} respectively) is computed based on theoretical considerations \cite{RN147, RN119, RN132} whereas in the DCP model it is estimated directly through a fitting of the experimental data \cite{RN114}. Furthermore, unlike the DL model, the contributions from the interband transitions to the imaginary part $\epsilon''$ of the dielectric permittivity are negative and unphysical for the DCP model \cite{RN114, RN125} (Figures S10 and S11). Regardless, this should have no actual bearing on the final results from the DCP model because the computation of the Mie resonances here ultimately depends only on the dielectric permittivity $\epsilon$. The results from the DCP model are therefore independent of the way the interband transitions are computed if the model itself is estimated based on accurate experimental data and approximates it closely.

\begin{figure}[ht!]	
\centering
\includegraphics[scale=0.64]{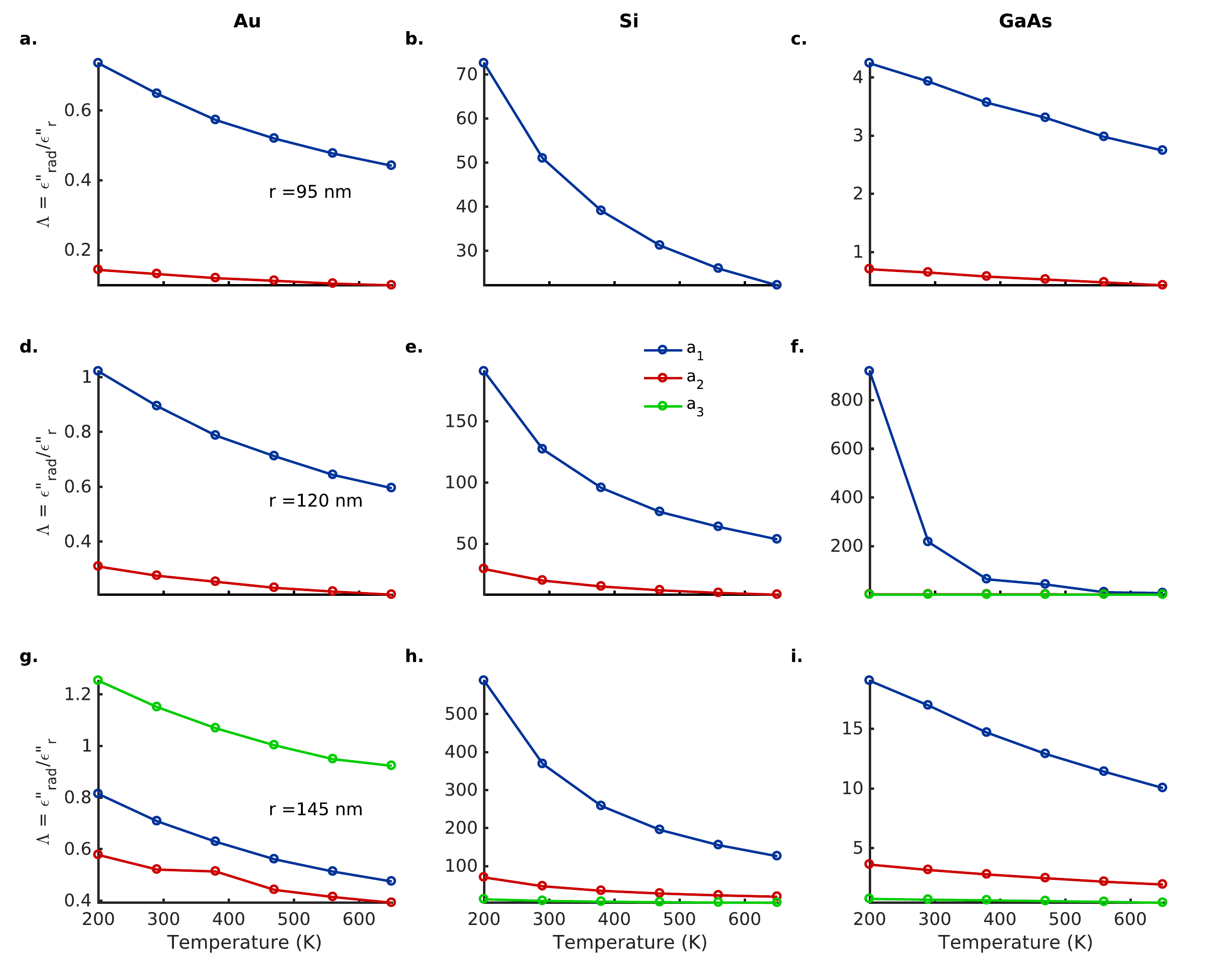} 
\caption{The ratio, $\Lambda$, of radiative ($\epsilon_\mathrm{rad}''(\lambda_n)$) to dissipative ($\epsilon_\mathrm{r}''(\lambda_n)$) damping for the strongest of the electric dipole (blue), quadrupole (red) and octupole (green) modes as a function of the temperature, $T$, for (a, d, g) Au (DL model), (b, e, h) Si and (c, f, i) GaAs nanoparticles of radii $r = 95, 120$ and $145$ nm, respectively. A decrease in $\Lambda$ is observed with rising temperatures for all materials. The Si and GaAs nanoparticles exhibit a $\Lambda$ that is about 1-2 orders of magnitude higher compared to the Au nanoparticles primarily due to the strong dissipative damping in Au. See Figure S12 for separate plots of $\epsilon_\mathrm{rad}''(\lambda_n)$ and $\epsilon_\mathrm{r}''(\lambda_n)$. Here, the Mie computations for the nanoparticles of different sizes take into account their thermal expansion, although the text labels indicate the values for nanoparticle radii at $200$ K.}
\label{rdamp} 
\end{figure}

\subsubsection*{Radiation Vs dissipative damping}
The dissipative damping of the $n^\mathrm{th}$ order resonance in a plasmonic nanoparticle is characterized by the imaginary part $\epsilon_\mathrm{r}''(\lambda_n) = \epsilon''(\lambda_n)/\epsilon_\mathrm{m}$ of the relative dielectric permittivity of its material and is therefore straightforward to estimate as a function of the temperature. However, an estimation of the radiation damping for a resonant Mie mode associated with a plasmon resonance presents some challenges. Experimentally, the radiation damping for a plasmon resonance is estimated by fitting a Lorentzian characterized by a linewidth $\Gamma_\mathrm{eff} (=\Gamma_\mathrm{D}+\Gamma_\mathrm{rad}$) to the extinction data obtained from a suspension of ideally monodisperse nanoparticles \cite{RN120, RN141}. This estimated linewidth $\Gamma_\mathrm{eff}$ represents the total damping associated with the plasmonic resonance. The radiation damping $\Gamma_\mathrm{rad} (= 2\hslash\xi V)$ itself is taken to be proportional to the volume $V$ of the particle with $\xi$ as its strength. The radiation damping strength $\xi$ is then estimated by a linear fitting of the total damping for the different volumes $V$ of the plasmonic nanoparticle with $\Gamma_\mathrm{D}$ and $\Gamma_\mathrm{rad}$ as free parameters in the linewidth, $\Gamma_\mathrm{eff}$ \cite{RN120, RN141}. The same procedure can also be employed if the extinction curves have been obtained from the finite difference time domain simulations of the plasmonic resonance  \cite{RN141}. This approach to determine $\Gamma_\mathrm{rad}$ however fails, if the plasmonic resonance occurs too close to an interband transition as is the case in our results for the Si and GaAs nanoparticles \cite{RN141}. 

On the theoretical side, one is again limited to using analytical results in the limit of a small but finite particle size. In such a case, the radiation damping ($\equiv \epsilon_\mathrm{rad}''(\lambda_n)$) for the $n^{\mathrm{th}}$ order Mie mode dominates over the dissipative damping characterized by $\epsilon_\mathrm{r}''(\lambda_n)$ when the following condition is satisfied  \cite{RN112}
\begin{equation}
\epsilon_\mathrm{r}''(\lambda_n) \ll \frac{x_{n}^{2n+1}}{n[(2n-1)!!]^2} \equiv \epsilon_\mathrm{rad}''(\lambda_n).
\label{ratiodamp}
\end{equation}
Figure \ref{rdamp} shows the ratio $\Lambda$ of the measure of the radiative damping  $\epsilon_\mathrm{rad}''$ to the imaginary part $\epsilon_\mathrm{r}'' (= \epsilon''/\epsilon_\mathrm{m})$ of the relative dielectric permittivity at the resonance wavelength ($\lambda_n$) as a function of the temperature for the strongest of the electric modes ($a_n$). Additionally, Figure S12 shows the plots for $\epsilon_\mathrm{rad}''(\lambda_n)$ and $\epsilon_\mathrm{r}''(\lambda_n)$ separately while Figure S13 presents the actual electric resonances $a_n$ at the temperatures $200$ and $650$ K. The above expression in equation \eqref{ratiodamp} for the radiation damping $\epsilon_\mathrm{rad}''(\lambda_n)$ is strictly valid only for $x \ll 1$. This condition is not satisfied by the resonance wavelengths $\lambda_n$ for the sizes ($r = 95, 120$ and $145$ nm) of the nanoparticles considered here. Nevertheless, equation \eqref{ratiodamp} does offer useful qualitative insights into the relative strengths of the radiative and dissipative damping in the nanoparticles. For the above reasons and in the absence of a better alternative, we employ $\Lambda (= \epsilon_\mathrm{rad}''/\epsilon_\mathrm{r}'')$ to understand the relative contributions of the radiative and dissipative dampings to the plasmonic resonances observed in the Au, Si and GaAs nanoparticles. 

It is clearly seen in Figure \ref{rdamp} that $\Lambda$ decreases with an increase in the temperature from $200$ to $650$ K for all materials. This is a direct consequence of an almost linear increase in dissipative damping, characterized by $\epsilon_\mathrm{r}''(\lambda_n)$, with increasing temperatures for all materials (Figure S12). Consistent with the experimental findings, a comparison of the plots in Figure \ref{rdamp} for the different radii shows that in general the radiation damping increases with an increase in the volume of the nanoparticles \cite{RN120, RN141}. Among all three materials, the metallic Au nanoparticles present the smallest values for $\Lambda$ pointing to the presence of a strong dissipative behavior as a consequence of the Drude damping of free electrons in the conduction band (Figure \ref{rdamp}a,d,g). This is also borne by the stronger dissipative damping $\epsilon_\mathrm{r}''(\lambda_1)$ in Au nanoparticles for the dipole mode $a_1$ that exhibits the broadest resonances (Figures S12-S13a,d,g). Figure S12 further shows that the dissipative damping in the Au nanoparticles is dominated by the electric dipole mode followed by the quadrupole and octupole modes whereas for the semiconducting Si and GaAs nanoparticles the order is completely inverted. In Si and GaAs nanoparticles (Figure \ref{rdamp}b-c,e-f,h-i respectively), the lowest values for $\Lambda$ are observed for the higher order quadrupole and octupole resonances that are strongly absorbing. These resonances occur at wavelengths wherein the dissipative damping ($\propto \epsilon_\mathrm{r}''$) arising from the photogenerated free charge carriers is high (Figure S12b-c,e-f,h-i respectively). The same holds true for the Au nanoparticles as well with the exception of the $145$ nm particles that exhibit a higher $\Lambda$ for the octupole resonance compared to the quadrupole and dipole modes. This is a consequence of the broad minima in $\epsilon_\mathrm{r}''$ (Figure S4d) observed around the resonance wavelength $\lambda_3 = 582$ nm for the electric octupole mode. In general, however the rapidly increasing denominator in equation \eqref{ratiodamp} with the order $n$ of the resonant mode implies that the radiation damping $\epsilon_\mathrm{rad}''$ decreases faster for the higher order Mie modes. Overall, $\Lambda$ is about one to two orders of magnitude stronger for the Si and GaAs nanoparticles compared to the Au nanoparticles (Figure \ref{rdamp}).

\section*{SUMMARY AND CONCLUSIONS}
Typically, depending on the target application, there arises a need to either maximize or minimize the dissipation of the absorbed energy through heat generation or scattering. Applications such as photothermal therapy \cite{RN31,RN32}, photocatalysis \cite{RN133}, high density magnetic memory devices \cite{RN136}, solar-thermophotovoltaics \cite{RN124} seek to maximize and harness the dissipative damping that drives the conversion of the incident radiation to heat. Conversely, thermal management through the development of high temperature insulation \cite{RN135}, cool reflective coatings \cite{RN46}, spectrally selective filters \cite{RN47} among others exploits the plasmonically enhanced scattering efficiencies resulting from a maximization of the radiative damping and a reduction of the dissipative damping to reduce conductive heat transfer. On the other hand, nanoscale cavity resonators used in sensing applications \cite{RN142} require a suppression of both dissipative and radiative damping mechanisms. It is therefore important to determine if a given material is suitable for a target application by understanding the temperature dependent absorption and scattering behavior in plasmonic materials. To this end, we have employed the experimentally estimated models of temperature dependent dielectric permittivity with Mie theory to model the thermoplasmonic behavior of Au (metallic), and, Si (indirect bandgap) and GaAs (direct bandgap) semiconductor nanoparticles. In addition, we have investigated the differences in the thermoplasmonic behavior of Au nanoparticles using the two widely employed Drude-Lorentz (DL) and the Drude and critical points (DCP) models of dielectric permittivity \cite{RN147, RN114, RN118, RN119, RN132}. 

Briefly, the real ($\epsilon'$) and imaginary ($\epsilon''$) parts of the dielectric permittivity for Au (DL model), Si and GaAs exhibit a monotonic increase with temperature. In contrast, their temperature dependence obtained from the DCP model for Au is complex and non-monotonic. Clear differences relating to the temperature dependence of the dielectric permittivity of Au exist between the DL and DCP models. An increase in $\epsilon'$ with temperature is observed for the DL model while the opposite occurs for the DCP model. As a result, this gives rise to a redshift in the scattering and absorption resonances computed with the DL model and a blueshift for the DCP model. The redshift of the Mie resonances with an increase in temperature is consistent with the experimentally determined temperature dependence of the plasmonic resonance for Au nanoparticles embedded in a silica matrix \cite{RN122}.  

A broadening and redshifting of the scattering and absorption resonances with an increase in the temperature and the particle radii is observed in nanoparticles of all three materials including Au (DL model), Si and GaAs. However, there is a massive Drude broadening of the Mie modes in the Au nanoparticles from free carrier absorption at longer wavelengths, away from the interband transitions. This broadening is enhanced at elevated temperatures as a result of the increased \textit{e-e} and \textit{e-$\phi$} interactions. The smaller ($r=20$ nm) of the Au nanoparticles exhibit a decrease in the scattering and absorption efficiencies with temperature for the dominant electric dipole modes due to an increase in the dissipative Drude damping. In the larger Au nanoparticles, the decrease in the Mie efficiencies with temperature gets compensated with contributions from the higher order Mie modes. 

In contrast, Fano resonances are seen in the scattering ($Q_\mathrm{sca}$) and absorption ($Q_\mathrm{abs}$) efficiencies for the Si and GaAs nanoparticles. These can be ascribed to the occurrence of sharp higher order electric ($a_n$) and magnetic ($b_n$) resonances against a background of the broad quadrupole or dipole modes. The strength of the scattering resonances with an increase in the temperature stays almost the same for the semiconducting Si and GaAs nanoparticles unlike a decrease for the metallic Au particles. In general, the Si and GaAs nanoparticles present higher values for $Q_\mathrm{sca}$ and $Q_\mathrm{abs}$ compared to those for  Au with the exception of the smaller 20 nm particles. However, the absorption efficiency farther away from the absorption band edge at longer wavelengths goes to zero for the Si and GaAs nanoparticles in contrast to the Au nanoparticles. In addition, the absorption band edge becomes progressively diffuse and redshifted for the Si and GaAs nanoparticles at higher temperatures.

The ratio of the radiative to dissipative damping, $\Lambda$, decreases with an increase in temperature for the nanoparticles of all three materials. Furthermore, $\Lambda$ is observed to be one to two orders of magnitude higher for the Si and GaAs nanoparticles at all temperatures compared to those for Au. This points to the presence of strong dissipative damping in Au at elevated temperatures. The ratio $\Lambda$ for all materials is additionally observed to be strongest for the dipole, quadrupole and octupole modes in that order. The only exception occurs for the larger $145$ nm Au particles wherein the octupole resonance presents the highest $\Lambda$ due to its occurrence at a broad minima in $\epsilon''$. In general, the dissipative damping in Si and GaAs particles is strongest for the electric octupole resonance followed by the quadrupole and dipole modes while the opposite holds true for the Au nanoparticles. 

In conclusion, we have investigated, analyzed and compared the thermoplasmonic behavior of Au, Si and GaAs nanoparticles to assess their feasibility for high-temperature applications. At elevated temperatures, the noble metals suffer from problems of poor thermophysical and chemical stability due to strong dissipative damping that results in excessive heating \cite{RN145}. Our results show that, unlike noble metal particles, semiconductor nanoparticles with their tunable plasmonic resonances, do not exhibit any significant deterioration in scattering and absorption efficiencies at high temperatures. Therefore, the use of semiconducting materials characterized by low dissipative damping is more suitable for applications that seek to exploit the plasmonic behavior of materials on the nanoscale under high temperature conditions.

\section*{ACKNOWLEDGMENTS}
The authors gratefully acknowledge funding and support from the Academy of Finland, COMP Center of Excellence Programs (2015-2017), Grant No. 284621; QTF Center of Excellence Program, Grant No. 312298; RADDESS Consortium Grant; the Aalto Energy Efficiency Research Program EXPECTS; the Aalto Science-IT project; the Discovery Grants and Canada Research Chairs Program of the Natural Sciences and Engineering Research Council (NSERC) of Canada; and Compute Canada (www.computecanada.ca).  

\section*{CONFLICT OF INTEREST}
The authors declare no conflict of interest.

Supplementary Information accompanies the manuscript on the Light: Science \& Applications website (http:/www.nature.com/lsa/).


\newpage

\begin{center}
\textbf{\large Supplementary Information: Thermoplasmonic response of semiconductor nanoparticles: A comparison with metals}

%
%

\end{center}
\setcounter{equation}{0}
\setcounter{figure}{0}
\setcounter{table}{0}
\setcounter{page}{1}
\renewcommand{\theequation}{S\arabic{equation}}
\renewcommand{\thefigure}{S\arabic{figure}}
\renewcommand{\bibnumfmt}[1]{[S#1]}
\renewcommand{\citenumfont}[1]{S#1}

\begin{figure}[ht!]	
\centering
\includegraphics[scale=0.72]{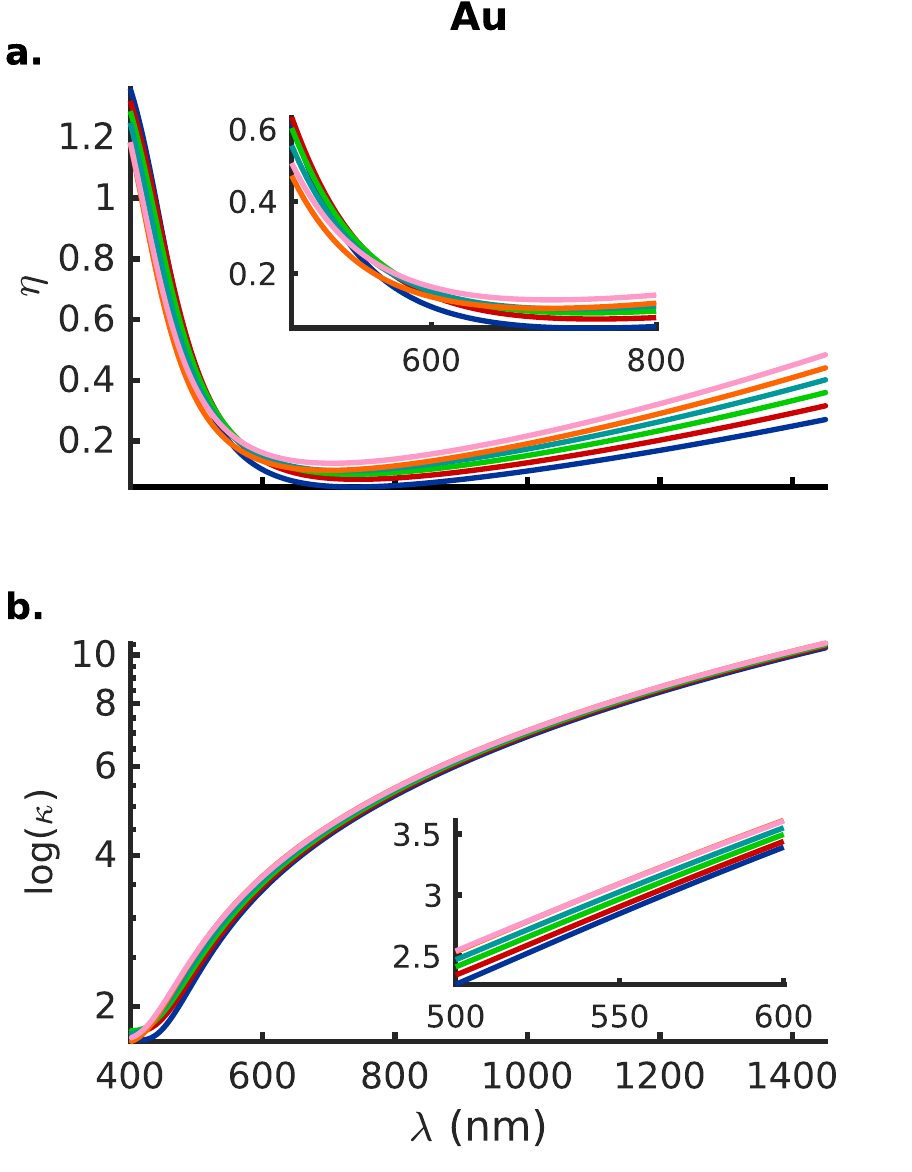}
\caption{The temperature dependent (a) real ($\eta$) and (d) imaginary ($\kappa$) parts of the refractive index of Au computed using the Drude and critical points (DCP) model. Unlike the Drude-Lorentz (DL) model (Figure 1a), $\eta$ in the DCP model behaves non-monotonically with temperature, $T$: $\eta$ transitions in the $500$-$740$ nm wavelength region from decreasing at shorter wavelengths to increasing at longer wavelengths beyond $\lambda = 740$ nm with rising temperature [see inset in (a)]. The values for $\kappa$ increase monotonically at all wavelengths (except for a small region between $\lambda = 400$-$425$ nm) for temperatures between $200$ and $560$ K and then saturate between $560$ and $650$ K [see inset in (b)].}
\label{Au-nk} 
\end{figure}

\begin{figure}[ht!]	
\centering
\includegraphics[scale=0.7]{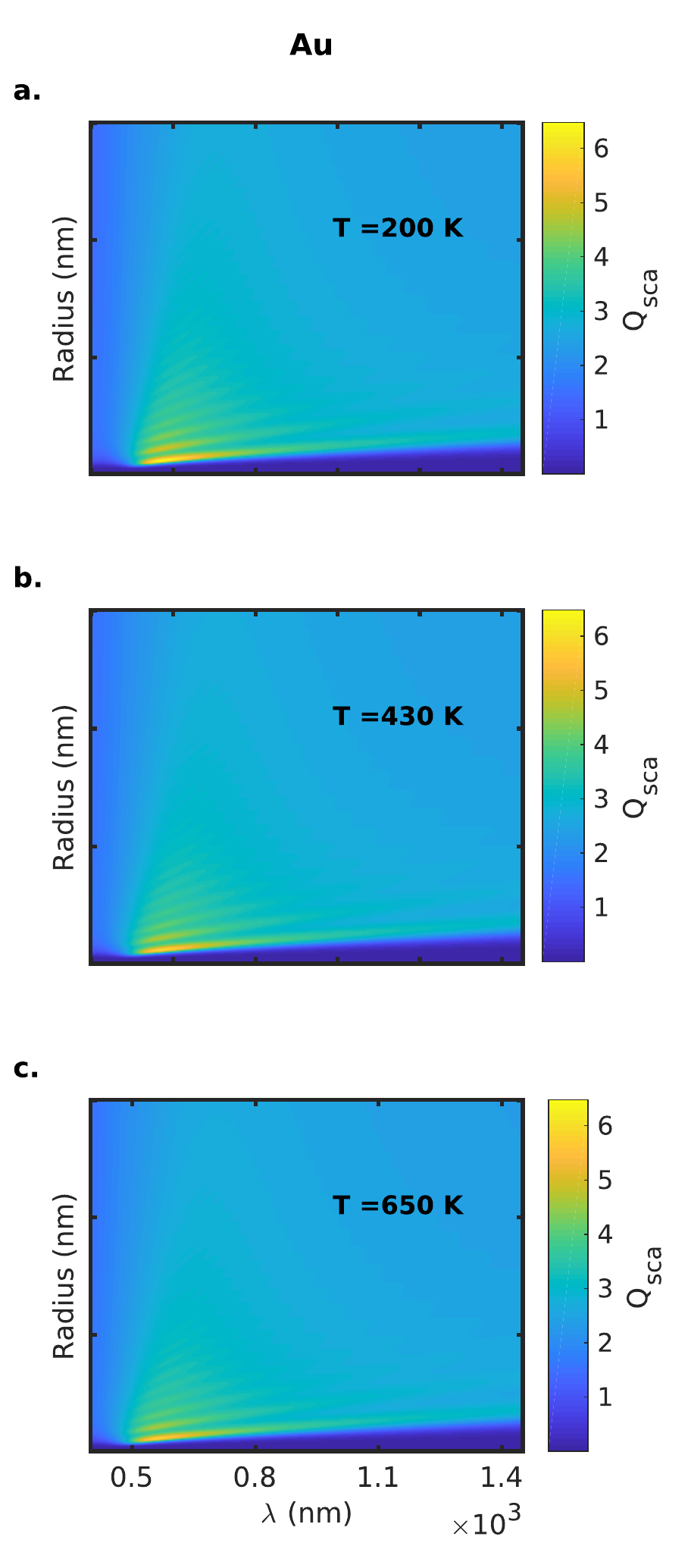}
\caption{Mie scattering efficiency, $Q_\mathrm{sca}$, computed using the DCP model for Au as a function of the wavelength, $\lambda$, of the incident radiation and the particle radii $r$ at three different temperatures, $T$, equal to (a) $200$, (b) $430$ and (c) $650$ K. Similar to the DL model, the strength of the resonances in $Q_\mathrm{sca}$ for Au particles show a decline with an increase in the temperature, $T$. See Figure \ref{Au-mQsca-DCP} for the clear blueshift observed in the scattering resonances for the DCP model as opposed to a redshift in the DL model (Figures 2a,d,g and 4a,d,g,j in the main text).}
\label{Au-Qsca2d} 
\end{figure}

\begin{figure}[ht!]	
\centering
\includegraphics[scale=0.7]{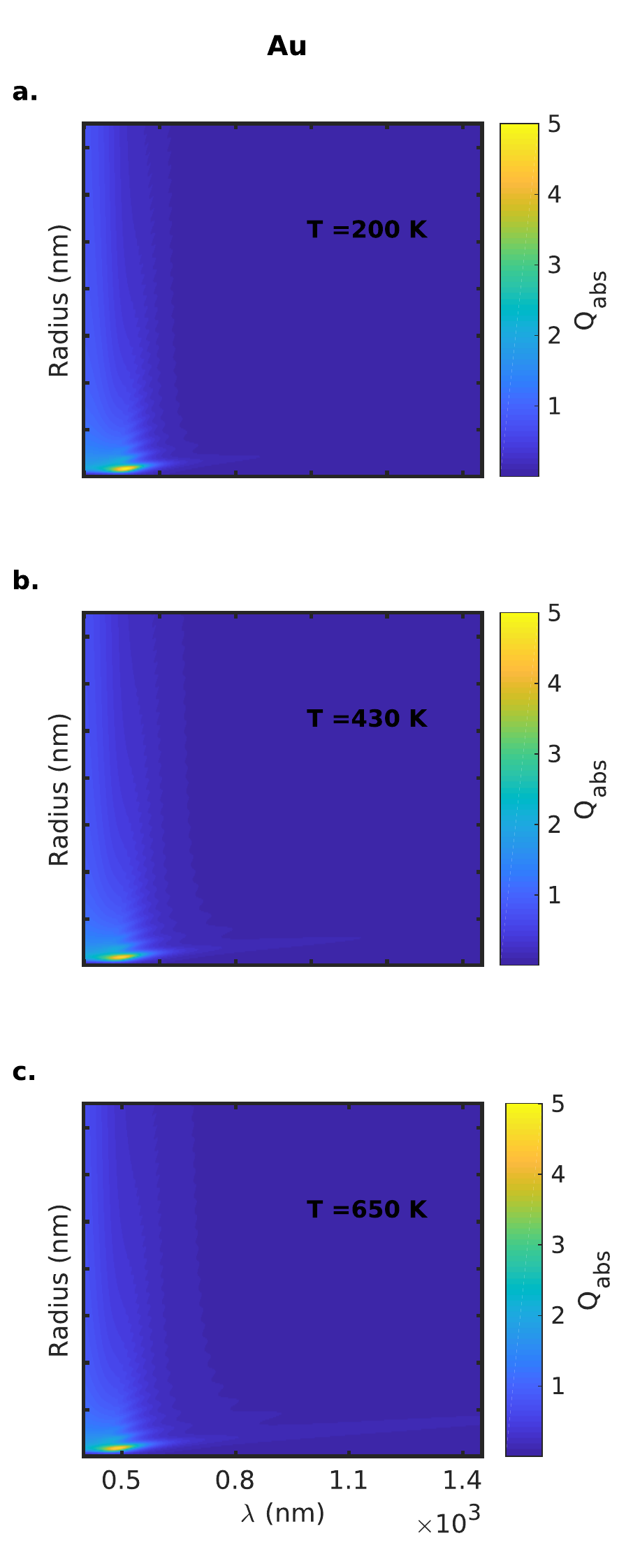}
\caption{Mie absorption efficiency, $Q_\mathrm{abs}$, computed using the DCP model for Au as a function of the wavelength, $\lambda$, of the incident radiation and the particle radii $r$ at three different temperatures, $T$, equal to (a) $200$, (b) $430$ and (c) $650$ K. See Figure \ref{Au-mQabs-DCP} for the clear blueshift is observed in the absorption resonances for the DCP model as opposed to a redshift in the DL model (Figures 3a,d,g and 7a,d,g,j in the main text).}
\label{Au-Qabs2d} 
\end{figure}

\begin{figure}[ht!]	
\centering
\includegraphics[scale=0.64]{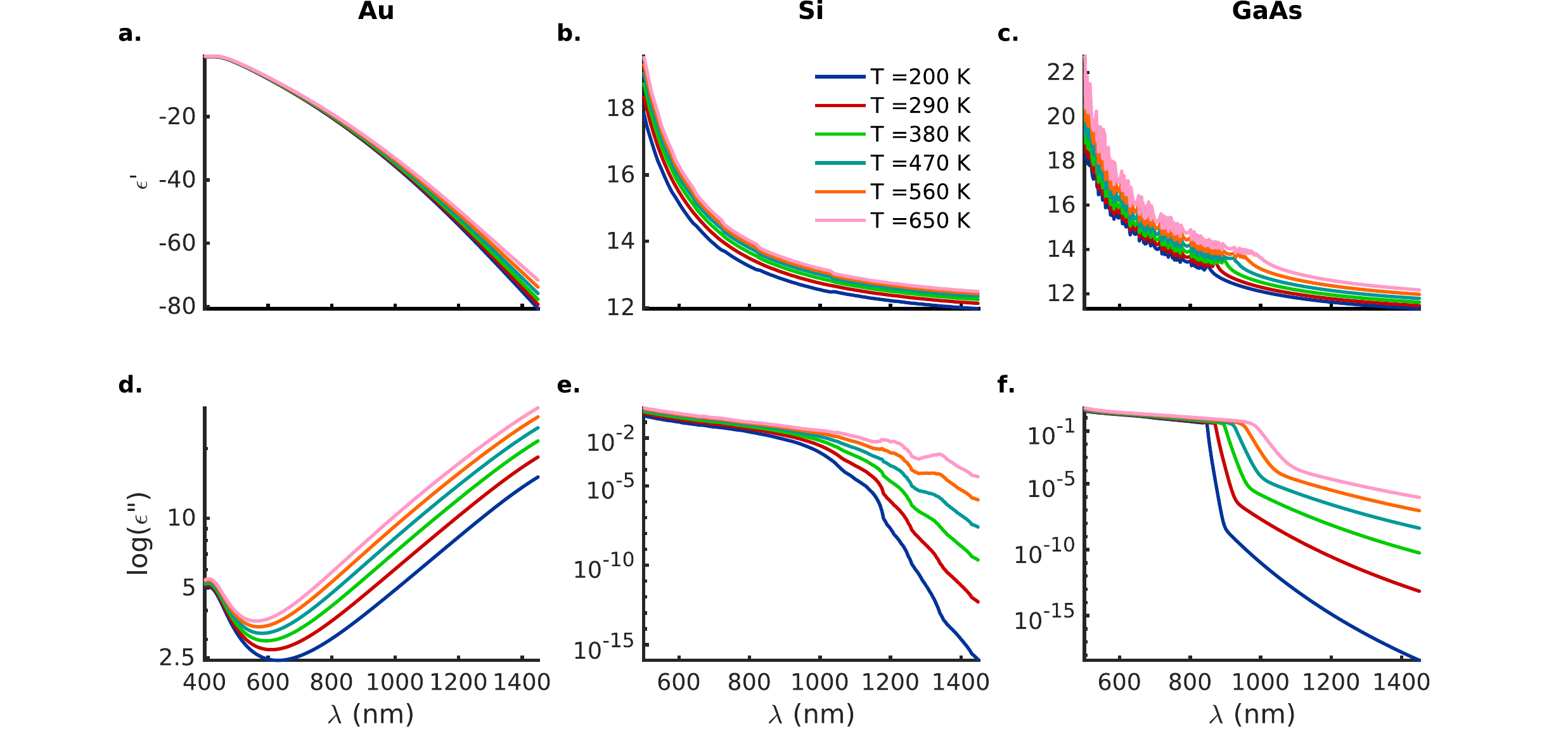}
\caption{Comparison of the temperature dependent (a-c) real ($\epsilon'$) and (d-f) imaginary ($\epsilon''$) parts of the dielectric permittivity for Au (DL model), Si and GaAs respectively. Both $\epsilon'$ and $\epsilon''$ are seen to increase monotonically with temperature, $T$, for all materials. Also, visible is the gradual rise in $\epsilon''$ for the indirect bandgap Si in contrast to the sharp rise observed in GaAs at shorter wavelengths beyond the onset of the absorption band edge. In (d), $\epsilon''$ for Au initially decreases with increasing wavelength and then increases monotonically for the longer wavelengths.}
\label{Au-DLeps} 
\end{figure}

\begin{figure}[ht!]	
\centering
\includegraphics[scale=0.64]{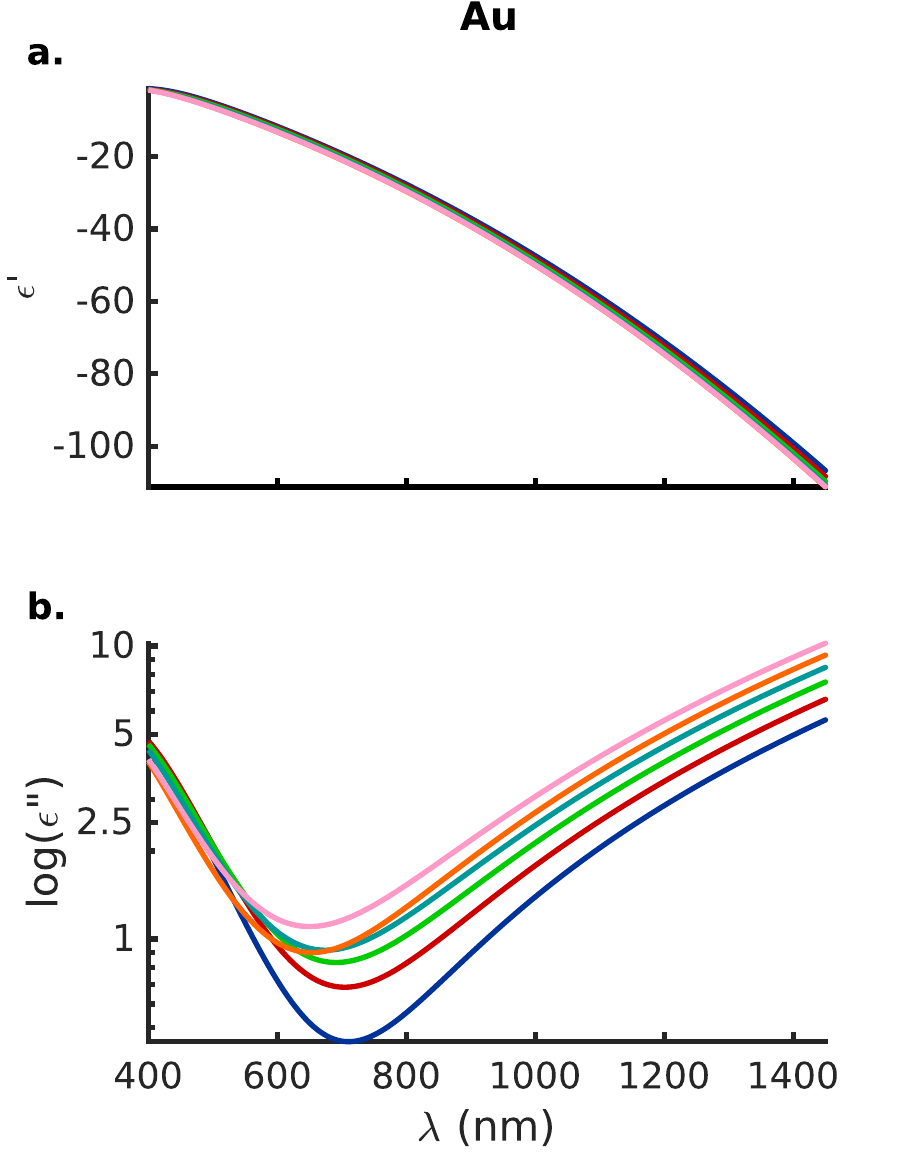}
\caption{The temperature dependent (a) real ($\epsilon'$) and (b) imaginary ($\epsilon''$) parts of the dielectric permittivity for Au computed using the DCP model. $\epsilon'$ displays a monotonic decrease with increasing temperatures across the entire wavelength range from $\lambda = 400$-$1450$ nm. On the other hand, the behavior of $\epsilon''$ is highly non-monotonic: it decreases with increasing temperature until $\lambda = 500$ nm followed by a transition region up to $\lambda = 710$ nm and an increase with temperature thereafter until $\lambda = 1450$ nm.}
\label{Au-epsDCP} 
\end{figure}

\begin{figure}[ht!]	
\centering
\includegraphics[scale=0.64]{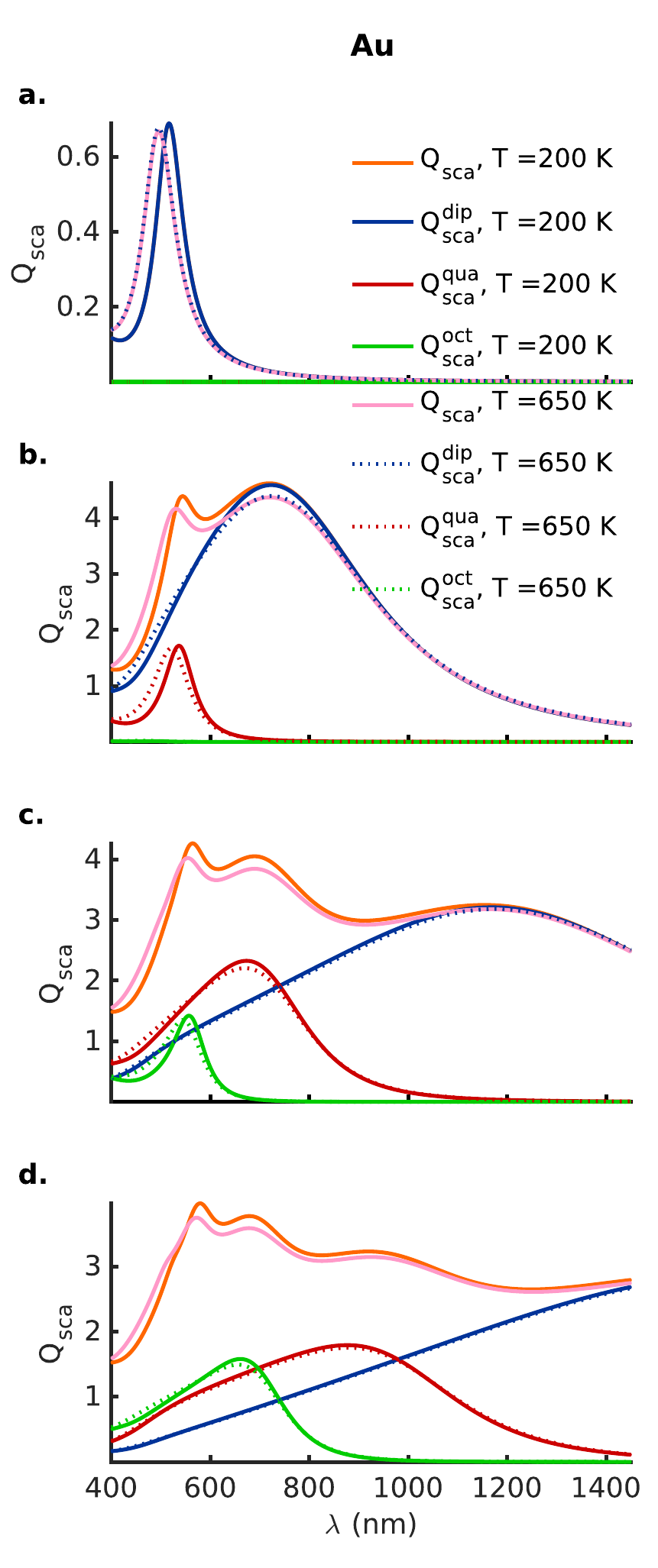}
\caption{Mie scattering efficiency ($Q_\mathrm{sca}$) with dipole (blue), quadrupole (red) and octupole (green) contributions as a function of the wavelength, $\lambda$, computed using the DCP model for Au nanoparticles of radii, $r$, equal to (a) $20$, (b) $80$, (c) $140$ and (d) $200$ nm at temperatures $T = 200$ (solid lines) and $650$ (dotted lines) K. The thick orange and pink solid lines represent the total scattering efficiencies, $Q_\mathrm{sca}$, at temperatures $T = 200$ and $650$ K, respectively. Just as in the DL model, the Mie resonances in the Au nanoparticles exhibit massive Drude broadening and a deterioration in strength with temperature increase due to dissipative damping. However, unlike the DL model (Figure 4a,d,g,j in the main text), there is a clear blue shift in the resonances with an increase in temperature. This blueshift is more pronounced for the smaller particles but gradually becomes weaker with increase in particle size. This is because the blueshift due to the decrease in the real part $\epsilon'$ of the dielectric permittivity at elevated temperatures is counteracted by the redshift due to plasmon retardation effect from an increase in the particle size. Here, the Mie computations for the nanoparticles of different sizes take into account their thermal expansion while the values indicated above for the nanoparticle radii are at $T = 200$ K.}
\label{Au-mQsca-DCP} 
\end{figure}

\begin{figure}[ht!]	
\centering
\includegraphics[scale=0.64]{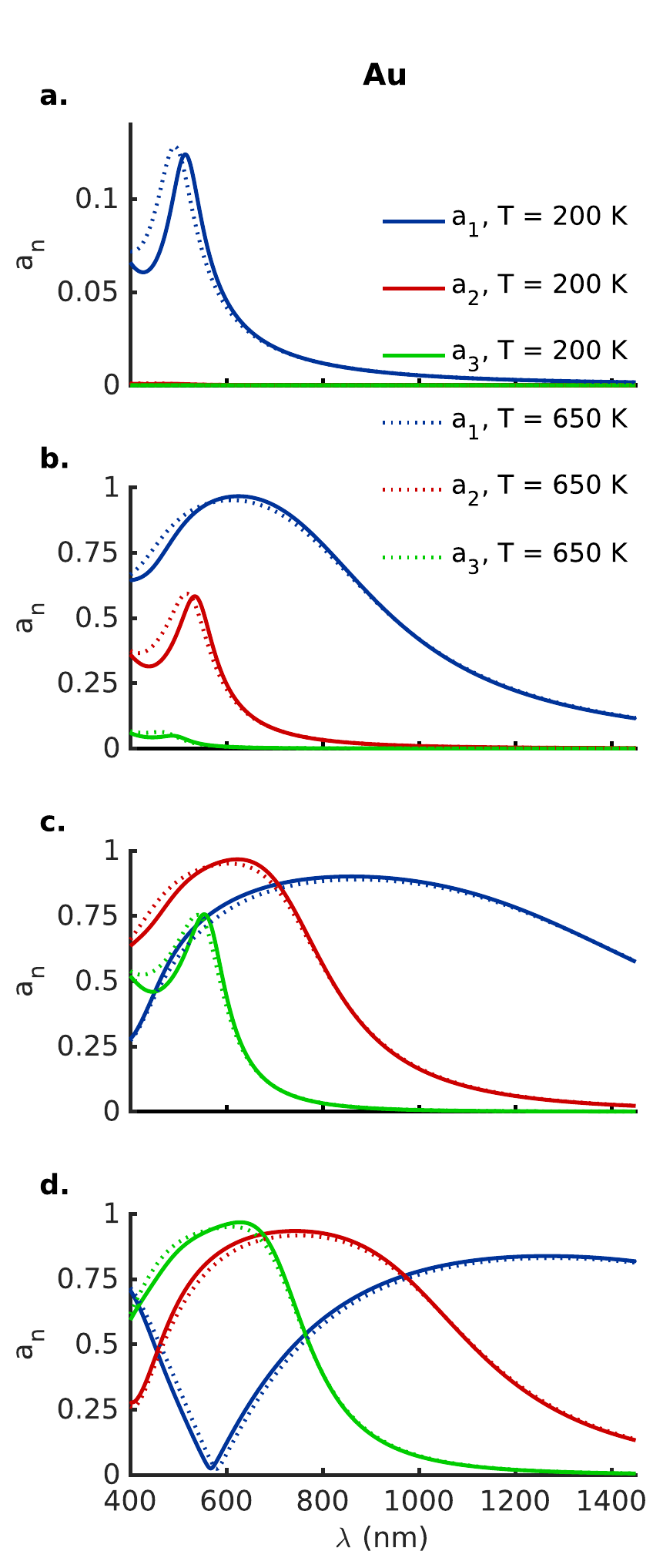}
\caption{Electric dipole (blue), quadrupole (red) and octupole (green) Mie modes ($a_n$) as a function of the wavelength, $\lambda$, computed using the DCP model for Au nanoparticles of radii, $r$, equal to (a) $20$, (b) $80$, (c) $140$ and (d) $200$ nm at temperatures $T = 200$ (solid lines) and $650$ (dotted lines) K. Unlike the DL model (Figure 5a,d,g,j in the main text), a blueshifting of the electric Mie resonances is observed with increasing particle size and temperature. This blueshift is more pronounced for the smaller particles but gradually becomes weaker with increase in particle size. This is because the blueshift, as a result of the decrease in the real part $\epsilon'$ of the dielectric permittivity at elevated temperatures, is counteracted by the redshift due to plasmon retardation effect from an increase in the particle size. Here, the Mie computations for the nanoparticles of different sizes take into account their thermal expansion while the values indicated above for the nanoparticle radii are at $T = 200$ K.}
\label{Au-An-DCP} 
\end{figure}

\begin{figure}[ht!]	
\centering
\includegraphics[scale=0.64]{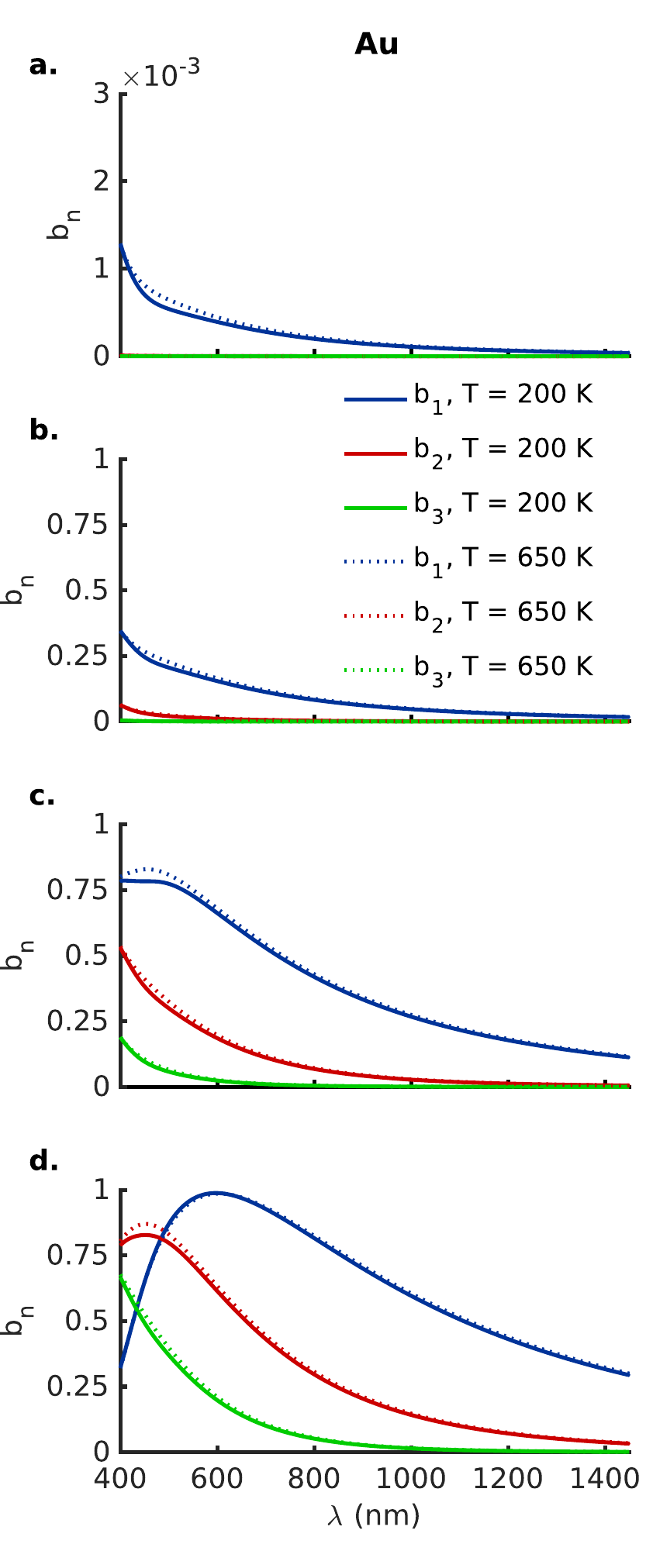}
\caption{Magnetic dipole (blue), quadrupole (red) and octupole (green) Mie modes ($b_n$) as a function of the wavelength, $\lambda$, computed using the DCP model for Au nanoparticles of radii, $r$, equal to (a) $20$, (b) $80$, (c) $140$ and (d) $200$ nm at temperatures $T = 200$ (solid lines) and $650$ (dotted lines) K. Similar to the electric Mie resonances $a_n$ (Figure \ref{Au-An-DCP}), the magnetic resonances also broaden with an increase in particle size and temperature for all materials. However, unlike the electric modes $a_n$ (Figure \ref{Au-An-DCP}), a blueshift of the magnetic Mie resonances is not observed with increasing temperature. Here, the Mie computations for the nanoparticles of different sizes take into account their thermal expansion while the values indicated above for the nanoparticle radii are at $T = 200$ K.}
\label{Au-Bn-DCP} 
\end{figure}

\begin{figure}[ht!]	
\centering
\includegraphics[scale=0.64]{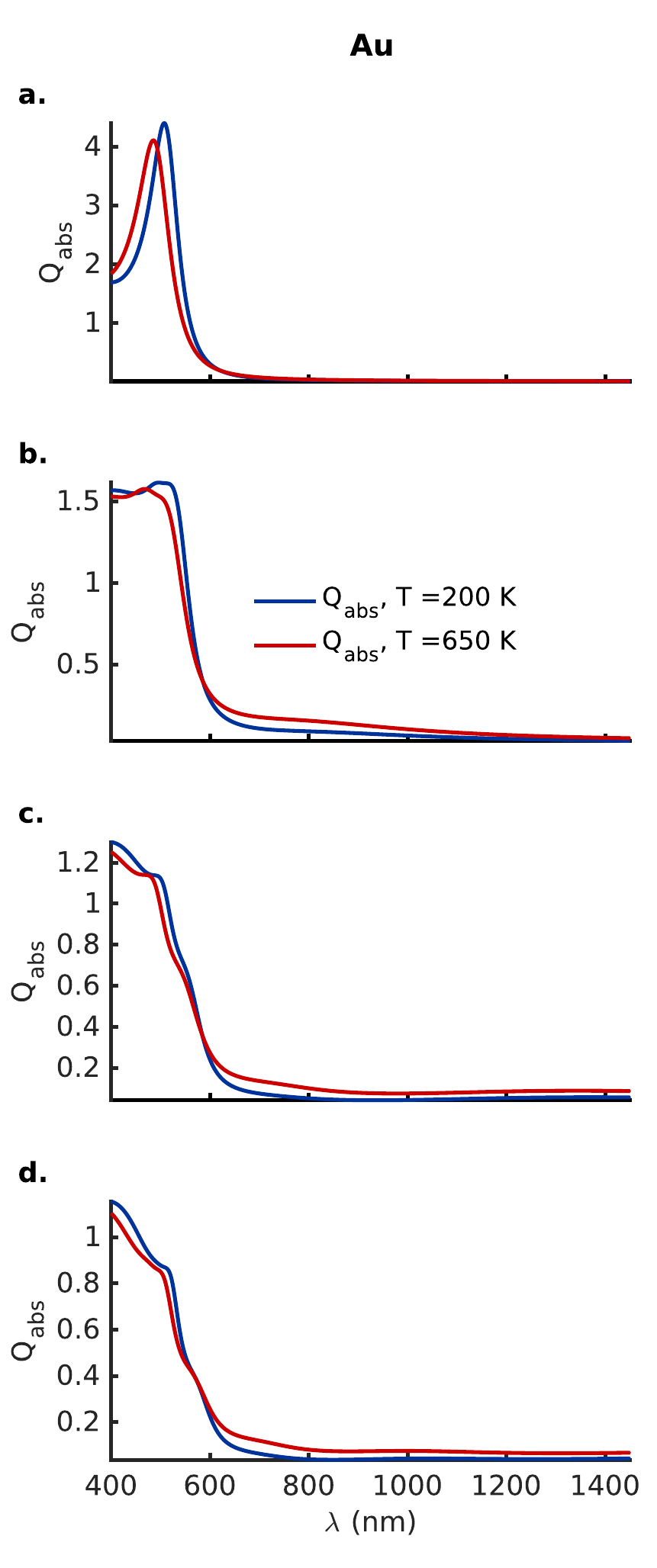}
\caption{Mie absorption efficiency ($Q_\mathrm{abs}$) as a function of the wavelength $\lambda$ computed using the DCP model for Au nanoparticles of radii, $r$, equal to (a) $20$, (b) $80$, (c) $140$ and (d) $200$ nm at temperatures $T = 200$ (blue) and $650$ (red) K. Similar to the DL model (Figure 7a,d,g,j in the main text), the highest $Q_\mathrm{abs}$ is observed for the smallest 20 nm particles. An increase in $Q_\mathrm{abs}$ with temperature for the larger Au nanoparticles is evident at the longer wavelengths due to uniform Drude broadening but the increase is not as pronounced as seen in the DL model (Figure 7a,d,g,j in the main text). There is also a blueshift of the absorption resonances primarily due to contributions from the blue shifted electric resonances $a_n$ (Figure \ref{Au-An-DCP}). Here, the Mie computations for the nanoparticles of different sizes take into account their thermal expansion while the values indicated above for the nanoparticle radii are at $T = 200$ K.}
\label{Au-mQabs-DCP} 
\end{figure}

\begin{figure}[ht!]	
\centering
\includegraphics[scale=0.64]{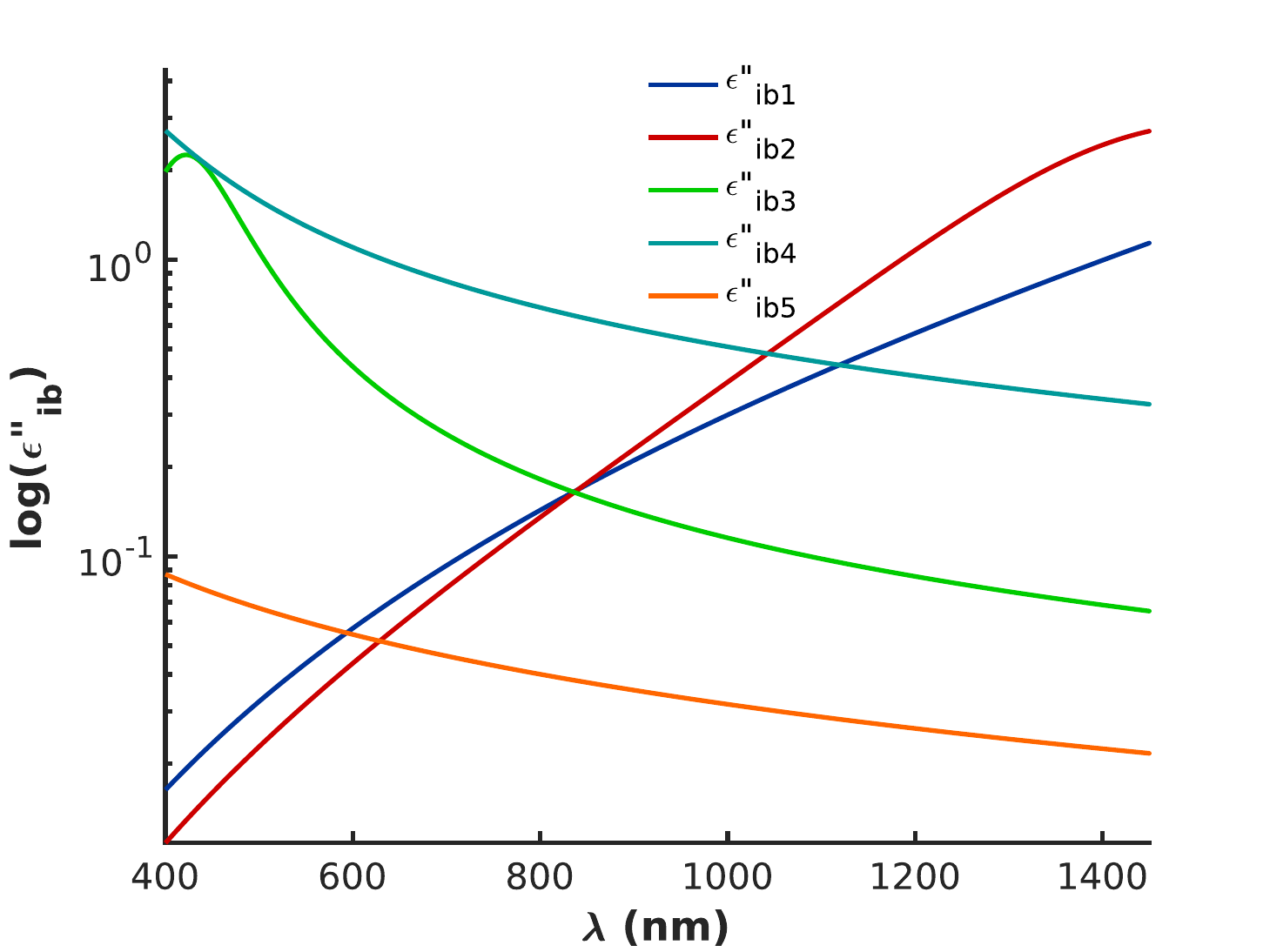}
\caption{The contribution of the temperature independent interband transitions, represented by the Lorentz oscillators, to the imaginary part $\epsilon''$ of the dielectric permittivity as a function of the wavelength, $\lambda$, of the incident radiation for Au in the DL model. The contributions from all of the five Lorentz oscillators stay positive throughout the wavelength range from $\lambda = 400$ to $1450$ nm.}
\label{ibepsDL} 
\end{figure}

\begin{figure}[ht!]	
\centering
\includegraphics[scale=0.64]{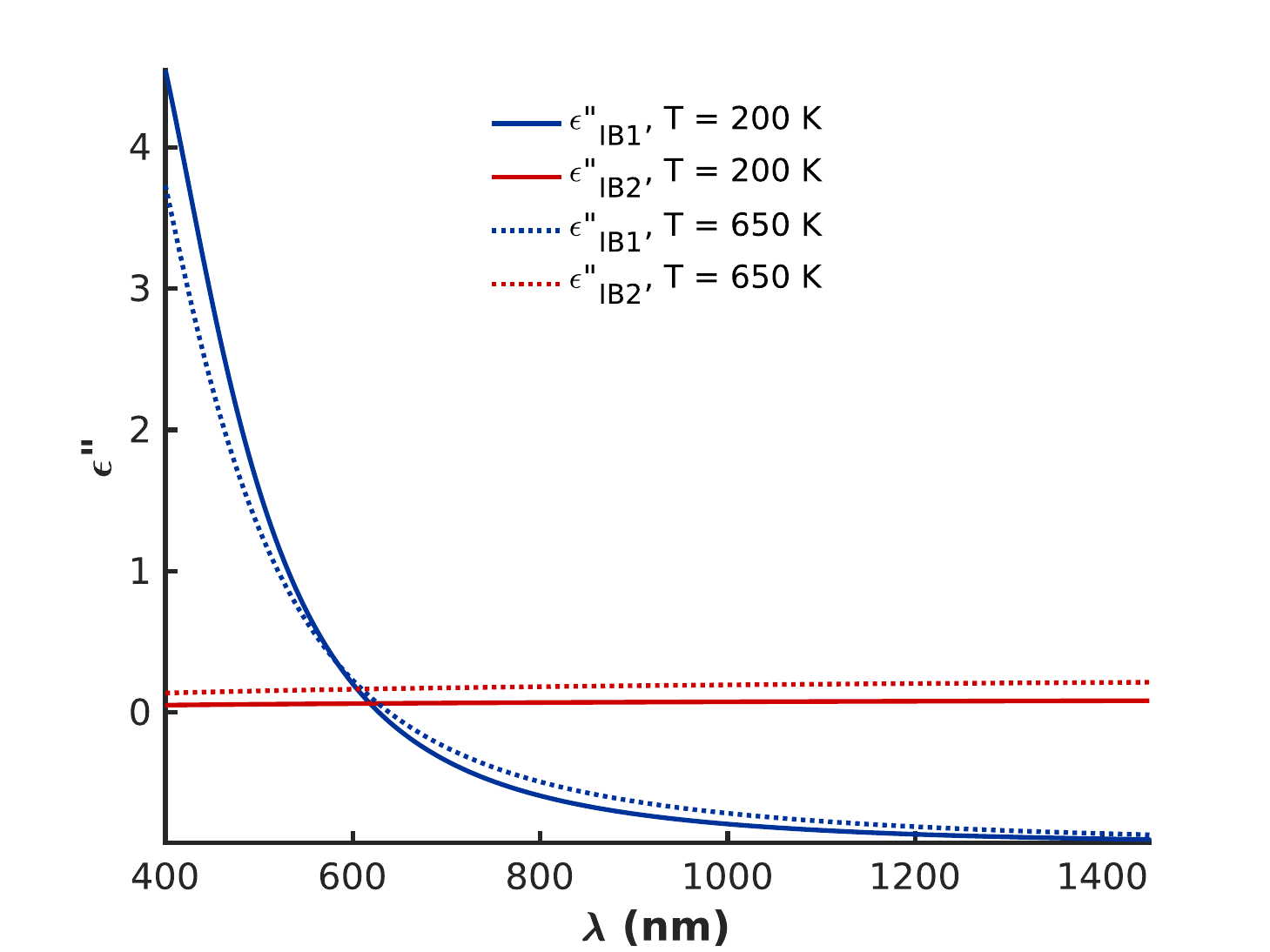}
\caption{Temperature dependence of the contribution of interband transitions to the imaginary part, $\epsilon''$, of the dielectric permittivity as a function of the wavelength, $\lambda$, of the incident radiation for Au in the DCP model. The contribution to dielectric permittivity from the interband transition occurring at $\lambda \approx 330$ nm in the DCP model becomes negative for $\lambda \gtrsim 600$ nm and hence is unphysical.}
\label{ibepsDCP} 
\end{figure}

\begin{figure}[ht!]	
\centering
\includegraphics[scale=0.64]{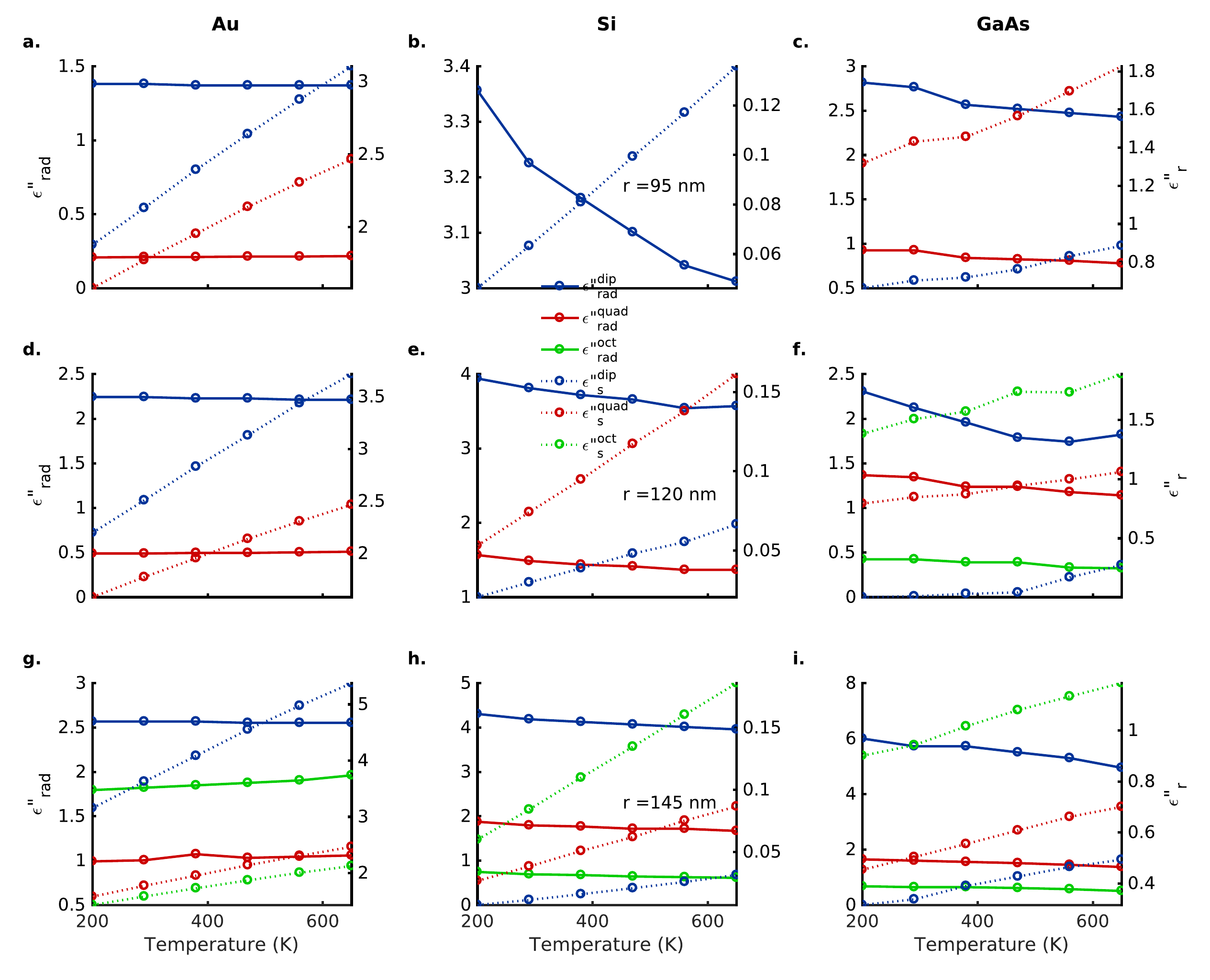}
\caption{The radiative ($\epsilon_\mathrm{rad}''(\lambda_n)$) and dissipative ($\epsilon_\mathrm{r}''(\lambda_n)$) damping for the strongest of the electric dipole (blue), quadrupole (red) and octupole (green) modes as a function of the temperature, $T$, for (a, d, g) Au (DL model), (b, e, h) Si and (c, f, i) GaAs nanoparticles of radii $r = 95, 120$ and $145$ nm, respectively. An almost linear increase in $\epsilon_\mathrm{r}''(\lambda_n)$ is observed with rising temperatures for all materials. The values of $\epsilon_\mathrm{r}''(\lambda_n)$ that govern dissipative damping are about $1$-$2$ orders of magnitude higher for Au nanoparticles compared to the Si and GaAs nanoparticles. This results in a $\Lambda (= \epsilon_\mathrm{rad}''/\epsilon_\mathrm{r}'')$ that is about $1$-$2$ orders of magnitude higher for the Si and GaAs nanoparticles compared to the Au nanoparticles (see Figure 8 in the main text). Additionally, consistent with experiments, the radiative damping is seen to increase with the particle size for all nanoparticles in general. Here, the Mie computations for the nanoparticles of different sizes take into account their thermal expansion, although the text labels indicate the values for nanoparticle radii at $200$ K.}
\label{compareRadDamp} 
\end{figure}

\begin{figure}[ht!]	
\centering
\includegraphics[scale=0.64]{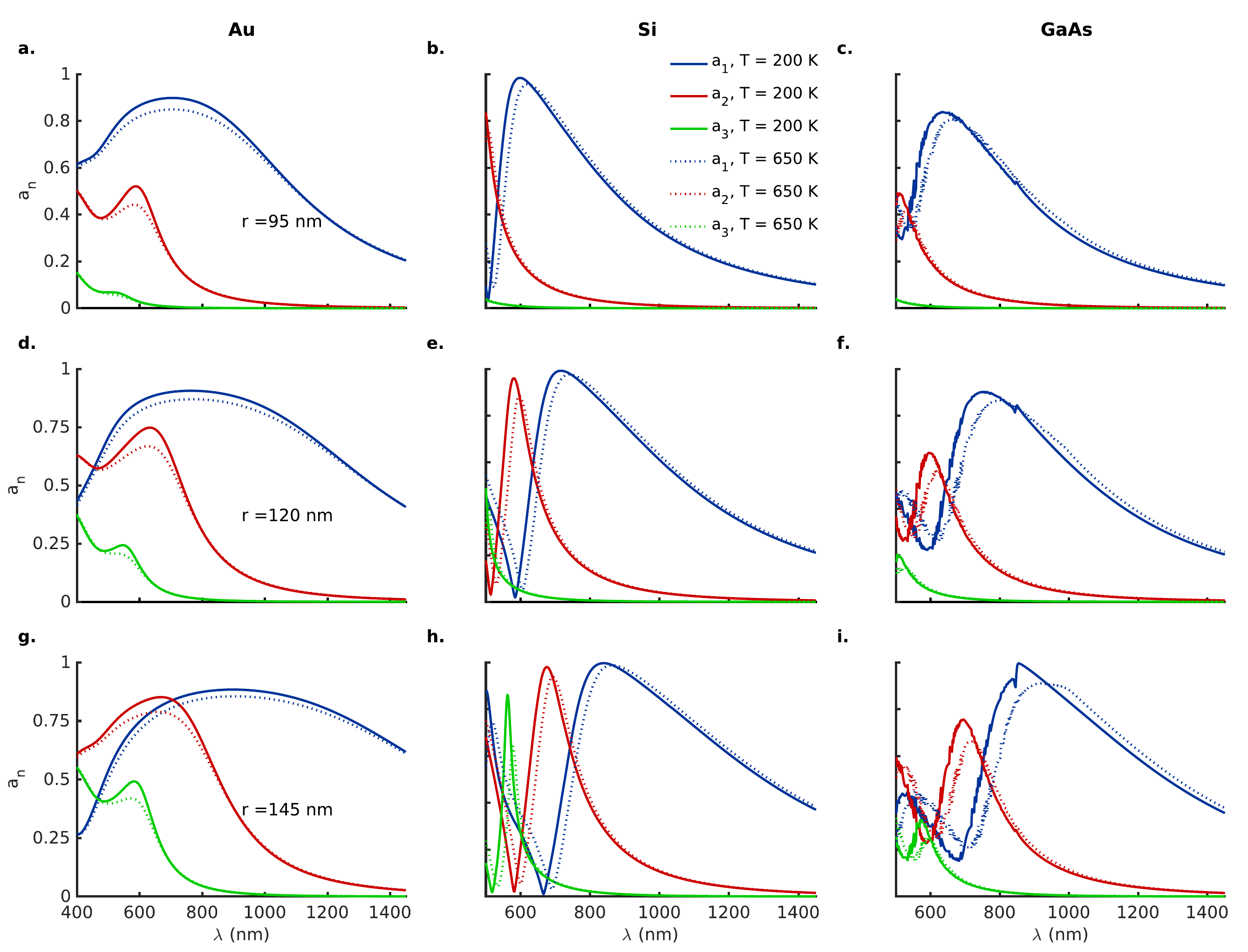}
\caption{Electric dipole (blue), quadrupole (red) and octupole (green) Mie modes ($a_n$) as a function of the wavelength, $\lambda$, for (a, d, g) Au (DL model), (b, e, h) Si and (c, f, i) GaAs nanoparticles of radii $r = 95, 120$ and $145$ nm, respectively, at temperatures $T = 200$ (solid lines) and $650$ (dotted lines) K. See Figure \ref{compareRadDamp} and Figure 8 in the main text for the corresponding radiative and dissipative damping alongwith their ratio, $\Lambda$, respectively. Similar to the electric modes $a_n$ in Figure 5 of the main text, a broadening and redshifting of the electric Mie resonances is observed with increasing particle size and temperature for all materials.}
\label{radDampAn} 
\end{figure}

\end{document}